\documentclass[sigconf]{acmart}
\AtBeginDocument{%
  }

\copyrightyear{2026}
\acmYear{2026}
\setcopyright{cc}
\setcctype{by}
\acmConference[IUI '26]{31st International Conference on Intelligent User Interfaces}{March 23--26, 2026}{Paphos, Cyprus}
\acmBooktitle{31st International Conference on Intelligent User Interfaces (IUI '26), March 23--26, 2026, Paphos, Cyprus}
\acmDOI{10.1145/3742413.3789140}
\acmISBN{979-8-4007-1984-4/2026/03}

\usepackage{graphicx}
\usepackage{subcaption}
\usepackage{booktabs}
\usepackage{xcolor}
\usepackage{caption}
\usepackage{booktabs}
\usepackage{cleveref}
\begin{document}

\title{Accepted with \emph{Minor} Revisions:\\Value of AI-Assisted Scientific Writing}

\author{Sanchaita Hazra}
\authornote{Both authors contributed equally to this research.}
\orcid{1234-5678-9012}
\affiliation{%
  \institution{The University of Utah}
  \city{Salt Lake City}
  \state{Utah}
  \country{USA}
}
\email{sanchaita.hazra@utah.edu.com}

\author{Doeun Lee}
\authornotemark[1]
\affiliation{%
  \institution{The Ohio State University}
  \city{Columbus}
  \state{Ohio}
  \country{USA}
  }
\email{lee.11501@osu.edu}

\author{Bodhisattwa Prasad Majumder}
\affiliation{%
  \institution{Allen Institute for AI}
  \city{Seattle}
  \state{Washington}
  \country{USA}
}
\email{bodhisattwam@allenai.org}

\author{Sachin Kumar}
\affiliation{%
 \institution{The Ohio State University}
 \city{Columbus}
  \state{Ohio}
 \country{USA}
 }
 \email{kumar.1145@osu.edu}

\renewcommand{\shortauthors}{Hazra et al.}
\newcommand{\sk}[1]{\textcolor{red}{[#1 -SK]}}
\newcommand{\skedit}[1]{\textcolor{red}{#1}}
\begin{abstract}
Large Language Models have seen expanding application across domains, yet their effectiveness as assistive tools for scientific writing—an endeavor requiring precision, multimodal synthesis, and domain expertise—remains insufficiently understood. We examine the potential of LLMs to support domain experts in scientific writing, with a focus on abstract composition. We design an incentivized randomized controlled trial with a hypothetical conference setup where participants with relevant expertise are split into an author and reviewer pool. Inspired by methods in behavioral science, our novel incentive structure encourages authors to edit the provided abstracts to an acceptable quality for a peer-reviewed submission. Our 2 x 2 between-subject design expands into two dimensions: the implicit source of the provided abstract and the disclosure of it. We find authors make most edits when editing human-written abstracts compared to AI-generated abstracts \textit{without} source attribution, often guided by higher perceived readability in AI generation. Upon disclosure of source information, the volume of edits converges in both source treatments. Reviewer decisions remain unaffected by the source of the abstract, but bear a significant correlation with the number of edits made. Careful stylistic edits, especially in the case of AI-generated abstracts, in the presence of source information, improve the chance of acceptance. We find that AI-generated abstracts hold potential to reach comparable levels of acceptability to human-written ones with minimal revision, and that perceptions of AI authorship, rather than objective quality, drive much of the observed editing behavior. Our findings reverberate the significance of source disclosure in collaborative scientific writing.
\end{abstract}

\begin{CCSXML}
<ccs2012>
   <concept>
       <concept_id>10003120.10003121.10011748</concept_id>
       <concept_desc>Human-centered computing~Empirical studies in HCI</concept_desc>
       <concept_significance>500</concept_significance>
       </concept>
 </ccs2012>
\end{CCSXML}

\ccsdesc[500]{Human-centered computing~Empirical studies in HCI}

\keywords{Generative AI, Human-AI Writing Assistance, Large Language Models, Scientific Writing, Text Editing and Evaluation, Alignment, Behavioral Science, Incentivized Random Trial}
% commented three lines below 

\maketitle

\section{Introduction}
% \paragraph{Paragraph}
Telling people about research is as important as doing it. Scientific publishing is an integral part of science, meant to disseminate research findings, foster collaboration, encourage reproducibility, and ensure that scientific knowledge is accessible and built upon over time~\citep{liang2024mapping}. Bad writing can and often does prevent or delay the publication of good science~\citep{gastel2022write}. Researchers and scientists today, however, are often not formally trained to write, and generally learn by imitating the styles of their advisors and other authors~\citep{lindsay2020scientific}. Large language models (LLMs), with their writing abilities, offer a promising venue to assist researchers in both conducting and writing about research. However, prior work on thorough assessment of their capabilities in open ended scientific writing tasks remains limited.

\begin{figure*}[t]
    \centering
    \includegraphics[trim=200 100 195 70,clip,width=0.9\linewidth]{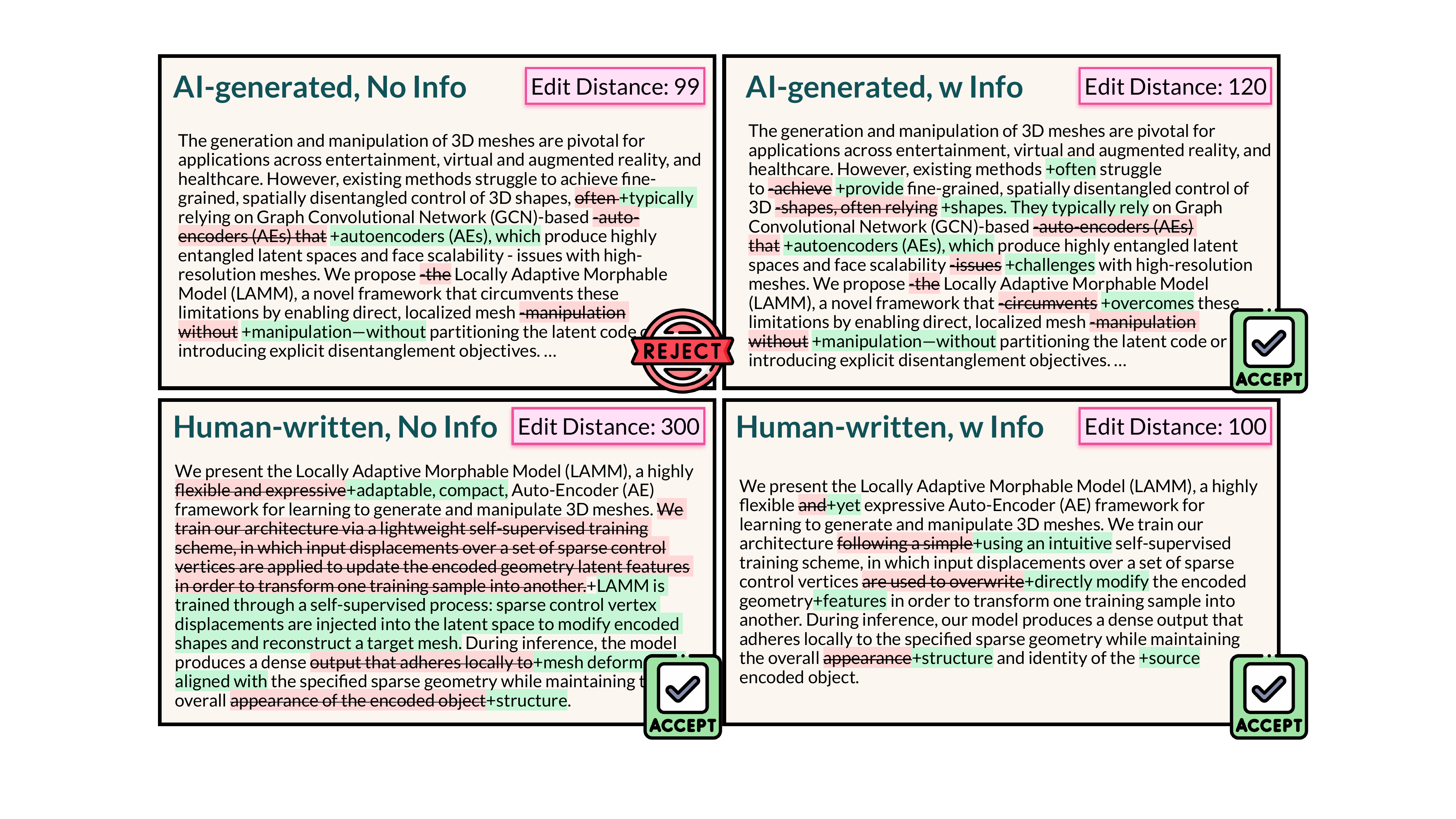}
    \caption{\textbf{Curious case of scientific writing}: We find that \textit{expert} authors (e.g., with PhD) make most edits when provided with human-written counterparts of AI-generated abstracts, especially so when the source of abstracts remains unattributed. With attribution, we see an opposite trend: authors made careful stylistic edits when the abstract was known to be AI-generated, which often raises the chance of getting the abstract accepted.}
    \label{fig:teaserfigure}
\end{figure*}

Even so, the use of language models continues to grow in this area. Many tools and platforms termed ``AI co-scientist'' have emerged in the past year \citep{gottweis2025towards,Beel2025EvaluatingSA}. They are marketed to be capable of generating, executing, and writing papers, with some of them getting accepted in academic workshops organized by peer-reviewed conferences \citep{yamada2025aiscientistv2workshoplevelautomated}. Popular scientific writing platforms such as Overleaf now offer writing assistant plugins.\footnote{\url{https://www.overleaf.com/learn/how-to/Writefull\_integration}} However, anecdotal examples of both seemingly AI-generated papers~\citep{cybernews_academic_2023,anthes_ai_2023} and peer reviews \citep{oransky_papers_2024} have inspired criticisms \citep{liang2024mapping}. In addition, across disciplines, recent studies have raised concerns over the declining quality of writing in scientific publications in several disciplines, attributing it to the rise of LLM-generated content~\citep{matsui2024delving,lemire2024will,geng2024chatgpt}. These studies, however, have largely focused on corpus-level trends on unreviewed manuscripts and are conducted without consulting or involving writers such as the authors of said publications. 

To raise awareness and prevent misuse by bad actors or unintentional use by over-reliant writers, there is an urgent need to conduct a systematic evaluation of the writing abilities of the current state-of-the-art models in specialized domains. Furthermore, despite current limitations, AI assistance holds tremendous potential in accelerating future scientific progress \citep{Sourati2023AcceleratingSW, Nathani2025MLGymAN, Majumder2024DatadrivenDW}. Towards understanding scientific writing capabilities, we seek to answer the following research question in this work: \textbf{maintaining the standard for scientific writing quality, to what extent do researchers rely on model-generated content for academic publishing?} 

To answer this question, we simulate a realistic writing environment where authors (domain experts), the intended users of an AI assistant, use and edit model-generated text to create scientific prose with the goal of getting it accepted by reviewers (also domain experts) for publication. To scope this work, we focus on the task of writing abstracts. An abstract is meant to summarize research findings, highlight key results, and pique the readers' interest. Given selected research findings, we evaluate how existing state-of-the-art LLMs perform at generating abstract drafts.

We simulate an abstract editing and reviewing process mimicking the process of scientific publication in a laboratory setting using the methodology of \textit{incentivized behavioral experiments}~\citep{azrieli2018incentives} to encourage participants to perform the task to the best of their ability. Authors edit the abstracts, and we quantify author reliance based on the extent of their modifications to make them submission-ready for a hypothetical conference publication. The task is conducted under two source information conditions: (1) authors \textbf{are informed} that the original abstract provided to them was AI-generated, and (2) authors \textbf{are not informed} that the abstract was assembled by generative AI. To contextualize this behavior with a more realistic workflow, we also provide the authors with human-written abstracts under the same source information conditions. By considering abstracts from different sources (human experts and LLMs), we assess the reception of LLM-generated abstracts by evaluating correlations between acceptance rates, edit frequency, and edit types.

The edited abstracts are then evaluated by a group of (incentivized) reviewers in a double-blind fashion, who vote to accept or reject the submitted abstract. This design allows us to examine how authors' reliance varies across source types and source information conditions, and how expert reviewers perceive the quality and credibility of abstracts produced through generative AI workflows. These experiments result in a collection of model-generated and human-edited texts, where we capture keystroke-level edits. From this rich data of inputs, abstracts, and edits, we conduct several evaluations and analyses. Our main evaluation is based on the correlation between the rate of editing and the final rating an abstract receives, contrasting the controls and different treatments. %where current models excel and where they could be improved
% .
We consider and measure the impact of variables like writer expertise (undergraduate students, graduate students, experienced researchers), and demographics and social factors (gender, perception of AI). 
In addition, we quantitatively (using linguistic tools) and qualitatively (via interviews with participants) characterize different kinds of edits authors make to understand the differences between edits made to AI- and human-authored abstracts. 

Across our experiments, we find that both the implicit origin of abstract and the disclosure of it systematically shape editing behavior and outcomes. Without source disclosure, authors make fewer edits to AI-generated abstracts, perceiving them as more readable, though PhD-level authors edited AI text more extensively. Despite these differences, reviewer evaluations showed no differences in accepting abstracts edited under no disclosure conditions. Disclosure of source primarily triggers social adjustments: authors reduced edits for human-written text when the source was revealed, whereas disclosure had only minor effects on AI-generated abstracts, with small impacts on perceived readability and confidence. 

Our quantitative analysis of stylistic metrics reveals that edits to AI-generated abstracts are to improve cohesion, reduce nominalizations, and produce informative sentences. On the contrary, authors tend to lengthen the openings of the human-written abstracts. Similarly, disclosure of source alone has a minimal direct effect on the writing style. Finally, thematic analysis of our interviews with authors confirms that authors adopt different strategies depending on the origin of the abstracts: they simplify AI-generated text while restructuring human-authored abstracts for emphasis.  Disclosure promotes a sense of accountability for both human- and AI-authored texts. Together, through a comprehensive large-scale experiment, we first highlight the varying cognitive editorial strategies emerging from writing that involves both human and AI-generated scientific writing.\footnote{Our experimental data, task instructions, and example of a task are provided as supplementary.}
% these results indicate that both text origin and authorship transparency shape not only the form but also the reasoning and cognitive strategies underlying scientific editing.
% \sk{edited based on new results, please verify}.
% s
% Our analysis reveals a significant shift in behavior when authors are presented with human-written abstracts without source attribution, leading to the highest level of editing. In contrast, when attribution is provided, authors make significantly more edits to AI-generated abstracts. Despite these differences in authorial effort, reviewer evaluations were not significantly influenced by the source of the abstract, indicating a potential disconnection between how authors revise and how reviewers assess final outputs. Edit distance is a significant negative predictor of final reviewer decisions, implying that extensive editing on either human-written or AI-generated abstracts does not necessarily improve acceptance chances. LLM-generated abstracts are often deemed sufficient in quality to be accepted with minimal intervention.

\section{Related Works}
\label{sec:related}

\subsection{AI Systems for Automating Scientific Research}
% AI-assisted writing tools have progressed from basic spell-checking and grammar correction to sophisticated language generation systems that can actively collaborate with human authors in scientific contexts \citep{Dhillon2024ShapingHC}. 
Early works on AI-assistance in scientific research largely focused on surface-level corrections and sentence-level improvements in writing \citep{Ito2019DiamondsIT,Bezerra2021NEWRITERAT} with some attempts at drafting entire paper structure \citep{wang-etal-2019-paperrobot}. Recent work has started to apply LLMs to the entire research cycle, including hypothesis discovery \citep{zhou2024hypothesis,zhang2025exploring,majumder2024discoverybench,abdel2025scientific}, experiment planning and execution \citep{li2025can,binbas2025mobllm,gui2025leveraging}, scientific writing \citep{kanna2024much,Liebling2025TowardsAA,Kousha2025HowMA}, and writing reviews \citep{silva2025ai}. Within the writing stage, some works have focused on specific subtasks like metaphor generation \citep{kim2023metaphorian}, figure caption generation \citep{kim2025multi}, citation generation \citep{xing-etal-2020-automatic,fierro2024learning,jung2022intent}, and related work generation \citep{https://doi.org/10.48550/arxiv.2212.09577,li2024related}. Such tasks are, typically,  accompanied by reference answers, making evaluation relatively straightforward. More recent work has started focusing on generating entire papers without any human intervention \citep{lu2024ai}. In this work, we adopt a more realistic setup of humans relying on an AI-assistant to generate drafts,  which they then edit \citep{kanna2024much}. We also use a principled method of human evaluation in the form peer review conducted via incentivized experiments. 

A wide range of commercial\footnote{\url{https://www.gatsbi.com/}, \url{https://anara.com/blog/ai-research-tools}} as well as open-source tools now offer services for every stage of the writing process, from literature review and synthesis to full manuscript drafting, editing, and refinement \citep{shao2025sciscigpt}. Specific features, such as automated abstract and title generators, are also becoming commonplace.\footnote{\url{https://www.writefull.com/}} The widespread availability and marketing of these tools signify a de facto adoption that necessitates rigorous empirical evaluation of their effectiveness, their impact on the quality of scientific output, and the human behaviors they engender, which our study addresses. In fact, large-scale corpus analyses suggest that AI assistance in scientific writing is already widespread \citep{kobak2025delving} in certain domains.
Recent studies emphasize the need to understand how human researchers interact with AI-generated drafts, how much agency they retain, and how these systems reshape scientific communication norms \citep{reza2025co}. Our work contributes to this line of inquiry by focusing on human reliance and editorial behavior when interacting with AI-generated scientific text under realistic publication incentives.

% LLMs based systems can support interactive writing assistance that can generate and rephrase text according to fine-grained author specifications, with studies showing author preference for AI-generated content in creative writing tasks \citep{Sun2021IGAAI}. 

\subsection{Human AI Collaboration and Co-writing}
The majority of prior research on human-AI co-writing has focused on creative writing \citep{singh2025systematic}. Using LLMs, researchers have developed and evaluated systems supporting story writing \citep{yuan2022wordcraft}, playwriting \citep{10.1145/3656650.3656688}, and character development \citep{park2025constella}. Some works also explore higher-level writing tasks such as prewriting \citep{wan2024felt}, and generating perspective-specific feedback \citep{rashkin2025help}.  Specialized creative applications have emerged for tasks including metaphor generation \cite{van2024perspective},
collaborative storytelling \citep{10.1145/3649921.3656987}, and personal diary writing \citep{10.1145/3613904.3642693}. User studies reveal complex dynamics in how writers interact with and perceive AI writing assistance. At a system interaction level, the design of AI suggestions significantly impacts user behavior and output: sentence-level suggestions promote original content creation, while paragraph-level suggestions improve efficiency \citep{10.1145/3544548.3581351,10.1145/3613904.3642134}. Writers' engagement with AI assistance is also influenced by
their personal values and goals \citep{10.1145/3532106.3533506,guo2025pen}. Writers show varying receptivity to AI support based on their
confidence levels, demonstrating higher acceptance in areas where they lack expertise \citep{al2025adoption,wang2025survey}, and
their desires for support are closely tied to their perception of support actors and personal values \citep{hwang202580,guo2024preserving}. Professional writers note persistent challenges with AI systems' ability to maintain consistent style and voice \citep{chakrabarty2024art}.

Moreover, this human-AI writing relationship raises important concerns. Studies reveal that biased AI models can influence not only the resulting text but also users' own opinions \citep{jakesch2023co}.
%These findings highlight a central tension: as AI writing systems become more sophisticated, they must balance providing assistance while preserving
% authenticity and agency. 
Prior research has identified biases, quality issues, lack of structure, and superficiality in AI-generated text, especially in domains like fiction and satire \citep{chakrabarty2024art, chakrabarty2024creativity, ippolito2022creative, marco2024pron,  mirowski2023co, mirowski2024robot}. Researchers have also found LLM-generated content to be overly positive and lacking in nuance and have highlighted the need for more diverse and representative training data to mitigate these shortcomings \citep{tian2024large}.
Our work adapts evaluation practices from prior studies in co-writing by focusing on the quality of AI-generated text using the amount of edits made to it by a human co-writer. Related to our work, experimental studies demonstrate that higher-level AI scaffolding can improve writing quality and efficiency, particularly for less experienced writers, while simultaneously raising concerns around diminished authorship and ownership \citep{10.1145/3613904.3642134}. In parallel, large-scale reviews of human–AI co-writing systems emphasize that writers desire different forms of AI support at different stages of the writing process, such as ideation, drafting, and revision \citep{reza2025co}. This underscores the importance of evaluating AI assistance within realistic workflows rather than isolated generation tasks. 
 
% \citet{lee2024design} emphasize that AI tools have significantly reshaped writing practices, introducing new benchmarks and expectations for the development of future AI-assisted writing systems. 

\subsection{Cognitive Biases in Human-AI interaction}
%papers related to our finding that if humans know the source of writing they edit differently.
The inability of humans to reliably detect AI-generated text makes their beliefs about a text's origin a powerful psychological variable, opening the door for cognitive biases to influence their judgment and behavior \citep{Jakesch2022HumanHF,henestrosa2024understanding,FIEDLER2025100321}. A rich literature in HCI and psychology documents two opposing, context-dependent human tendencies when interacting with automated systems \citep{844354,lee2004trust,DZINDOLET2003697}. On one hand, \textbf{automation bias} describes the tendency to over-rely on or excessively trust automated outputs, often using them as a cognitive shortcut to reduce mental effort \citep{mosier2018human,lyell2017automation}. This can lead to errors of commission (accepting incorrect AI-generated information) and errors of omission (failing to notice problems that the AI missed) \citep{ABDELWANIS2024460}.

On the other hand, \textbf{algorithm aversion} describes a tendency to prefer human judgment over algorithmic judgment, even in cases where the algorithm is demonstrably better or equivalent \citep{Sunstein_Gaffe_2025}. This aversion is often heightened in high-stakes, subjective, or ethically charged tasks, such as medical diagnosis \citep{filiz2023extent}. It can be driven by a fundamental desire for human agency, a negative emotional reaction to being judged by a machine, and a belief that human experts possess unique, ineffable knowledge that algorithms cannot capture. 

Our study advances the literature by shifting the focus from perceptual measures of AI-generated text such as self-reported ``credibility'' or ``trust'' \citep{LERMANNHENESTROSA2023107445,MaC3019} to concrete, measurable behavior in terms of amount of edits made to the text. The finding that attribution significantly impacts the number of edits demonstrates that this cognitive bias translates into tangible, effortful action on the part of the expert, also mirrored by recent empirical work \citep{zhu2025humanbiasfaceai}. Controlled experiments in German theses by \citet{FIEDLER2025100321} show that when identical texts are labeled as AI-generated versus human-written, evaluators systematically favor the human-labeled versions, even when the labels are incorrect, demonstrating a strong source-identity bias against AI authorship. Conversely, when source information is withheld, both expert and non-expert evaluators struggle to reliably distinguish AI-generated academic writing from human writing, often rating them as comparable in quality.

\subsection{Peer Review and Scholarly Evaluation}
Our experimental design is situated within the broader field of science of peer review, which treats scholarly evaluation not as a purely objective process, but as a human endeavor subject to cognitive and social biases. A significant body of literature has used experimental methods to uncover these biases. For instance, landmark studies have shown that reviewer recommendations are significantly influenced by the prestige of an author's institution, a bias that double-blinding is intended to mitigate \citep{ceci1982peer,tomkins2017single}. Relatedly, we study the perceived origin of the text itself as a bias. The attribution of an abstract to either a human or an AI serves as an analogue to author identity, allowing us to investigate whether a ``source identity bias'' influences the authorial and review process in a similar manner to established forms of author bias.

% Designing one-shot benchmarks for evaluating scientific writing capabilities of models that reflect their real-world use is hence infeasible and impractical with annotators with no incentive or experience in scientific writing. 
In this work, following the methodology of behavioral scientists, we conduct incentivized randomized controlled trials of increasing complexity for this evaluation. Randomized controlled trials are widely regarded as the gold standard for establishing causal relationships in behavioral interventions \citep{banerjee2009experimental}. Monetary incentives promote task engagement and data reliability. Participants align the study's rewards with their decisions through a cost-benefit analysis, yielding behavior that closely mirrors real-world decision-making \citep{gneezy2000pay}. Recent studies have employed RCTs to assess the effectiveness of AI tools in various domains, including writing and education \citep{kizilcec2015attrition}. These trials provide empirical evidence on how AI interventions influence user outcomes, supporting their utility in scientific research.

\begin{figure*}
    \centering
    \includegraphics[trim=10 400 15 15,clip,width=\linewidth]{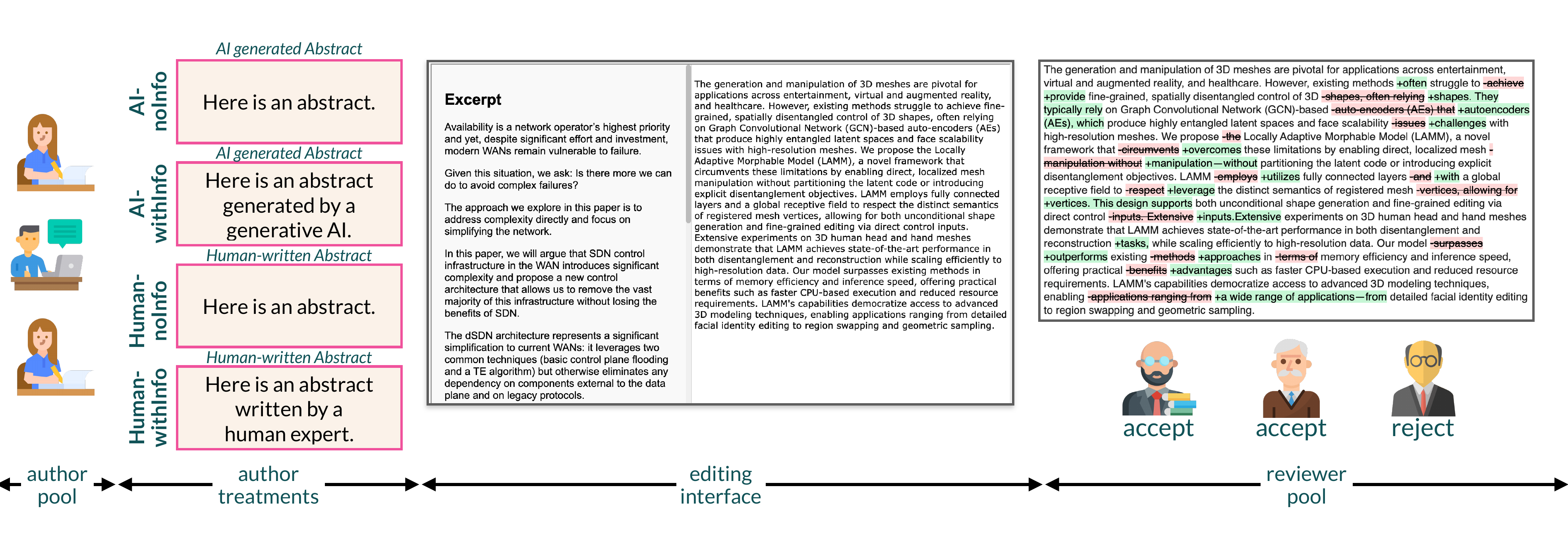}
    \vspace{-2em}
    \caption{\textbf{Experimental workflow}: Authors from a shared pool were randomly assigned to one of four treatments varying abstract source (AI-generated vs. human-written) and information disclosure (with vs. without source information). Authors revised abstracts in the editing interface. Each edited abstract was randomly assigned to three reviewers for their individual verdict. Reviewers only see the final edited version of the abstracts. A majority vote decides the final verdict.}
    \label{fig:design}
\end{figure*}

\section{Experiment Design}

% ----------------
Writing scientific prose is a challenging task. It is intended to communicate complex ideas to the academic community or to the public, focusing on preciseness, clarity, and brevity~\citep{lindsay2020scientific}. It requires the synthesis of diverse specialized knowledge and information sources, including non-textual elements like tables, statistics, and figures, alongside long contexts with strict formatting requirements. Performed by domain experts such as researchers with varying expertise, it is an inherently iterative process \citep{mao2019data, cornell2021please}. It is often collaborative, involving people with different experiences contributing, editing, and commenting on each other's written work. 

We aim to create a test-bed to evaluate how humans perceive and evaluate scientific writing. We follow a 4-step approach:
(1) We select original research publications from peer-reviewed conference publication venues and curate AI-generated abstracts from the content of these papers using state-of-the-art LLMs;
(2) We engineer an incentivized hypothetical conference set-up where we recruit human participants online to form an Author and a Reviewer pool. This allows us to emulate a real-world writing setting for both parties to play their part in the study;
(3) We develop a web platform based on FirePad\footnote{The editing platform codebase is available at: \url{https://github.com/skai-research/scientific-writing-assistance}} to capture fine-grained character-level edits such as insertion, deletion, or substitution with timestamps. Each abstract is shown to the authors on this platform to generate edited abstracts. 
(4) We show each edited abstract to three independent reviewers to obtain the final accept/reject decision by majority voting. 

% \begin{table}

\subsection{Collecting and Curating Abstracts} \label{sec:abstractselection}
For the scope of this study, we focus on the computer science (CS) domain for several reasons. First, CS is a fast-paced field with high publication volume \citep{CVPR2024}, rapid turnaround cycles, and a high impact factor, particularly in conferences \citep{reuters2024aivstaylor, freyne2010relative}. This makes it a relevant and timely setting to study the integration of AI in scientific writing. Second, the widespread adoption of AI tools by researchers in CS provides ecological validity for examining human-AI collaboration in this context.
Finally, as CS researchers ourselves, we are well-positioned to assess the accuracy and relevance of domain-specific content. While our study centers on CS, the design is generalizable and can be extended to other scientific domains. A typical research process 
% in Computer Science 
involves identifying a research problem, reviewing existing literature, formulating a hypothesis, collecting data, analyzing the data, and potentially modifying the hypothesis and repeating the process. At the end, the researchers draw conclusions and present the findings. We 
% aim to understand 
evaluate the role of AI-assistance in the final stage, where, given research findings, \emph{an LLM generates a coherent abstract summarizing the results of the study in a factual, coherent, and appealing way.}

% \begin{wraptable}[12]{r}{0.4\textwidth}
\begin{table}
  \centering
  \small
\vspace{-1em}
\caption{Abstracts from papers across venues}
\begin{tabular}{lccc}
\toprule
Venue & Count & Average Citation & Year \\
\midrule
NeurIPS & 6 & 18.83 & 2024 \\
CVPR & 6 & 24.17 & 2024 \\
ACL & 6 & 30.33 & 2024 \\
SIGCOMM & 6 & 5.83 & 2024 \\
VLDB & 5 & 2.60 & 2024 \\
IEEE S\&P & 6 & 19.83 & 2024 \\
ICSE & 5 & 10.80 & 2024 \\
ASPLOS & 5 & 15.40 & 2024 \\
\bottomrule
\label{tab:venuelist}
\end{tabular}
% \vspace{-1em}
\end{table}
The most realistic setting to conduct our experiment is to involve different researchers who are at the end of their studies and wish to write up their findings. This setting, however, is expensive and prohibitively slow to emulate. Instead, we choose already published recent papers and convert them into a format of ``research excerpts'' that could be fed to an LLM to generate abstracts. We deliberately choose already published papers to isolate the effect of the quality of writing from the quality of the research itself; we only aim to evaluate the latter. 

We select 45 published papers from top-tier conference venues in CS.\footnote{To ensure that our selected research papers were not a part of the training dataset of the state-of-the-art LLMs, we use a subset of only very recent papers published in 2024 in the top CS conferences. For details on selected papers, please refer to \autoref{tab:paperlist}.} We mainly focus on application-based papers from subfields of CS, such as NLP, HCI, Systems, and others.\footnote{We refrain from using lengthy papers with theoretical underpinnings, as their heavy dependence on background knowledge may make it challenging to generate concise research excerpts without losing information.}
For each venue's latest publication, we randomly sample papers as shown in \autoref{tab:venuelist}. Using GPT-4o, we create research \textit{\textbf{excerpts}}, i.e., extract all relevant text from the paper that is necessary to write an abstract.\footnote{OpenAI’s GPT-4o stood as the most recent model available during the time of the experiment. Our conclusions are conditional on this temporal choice and do not presume continued popularity or dominance of the model or its provider.} We provide the exact prompt we use in \autoref{fig:promptforfindings} (Appendix). Note, we focus on extraction rather than generation to make sure no facts in the original paper are misrepresented in the generated plan. We also manually verify each model output for factual correctness. We provide a sample output in \autoref{fig:sampleresearchfindings} (Appendix).

Using these excerpts, we also generate versions of AI-generated abstracts with respect to their original counterpart, also using GPT-4o.\footnote{At the time of conducting this experiment, ChatGPT's GPT-4o was OpenAI's premier offering, with traffic surpassing 100 million visits upon its release \citep{pcmag_gpt4o_traffic_2024,duarteuserstatistics2025}. Since its inception, ChatGPT has been in use for writing abstracts, generating and editing text, and academic proofreading \citep{howard2024characterizing, al2024navigating, brameier2023artificial}.} We provide the prompt to generate the abstract and a sample abstract in \autoref{fig:promptforabstract} and \autoref{fig:sampleabstract} (Appendix). Corresponding original abstract from paper is provided in \autoref{fig:sampleabstract_human} (Appendix). 

In this paper, we refer to four versions of abstracts which we use accordingly henceforth. (1) We refer to the authentic human-written abstracts of the published research papers from the peer-reviewed conference publication venues that we used in our study as \textbf{original abstracts};
(2) We refer to the populated AI-generated versions of original abstracts of the respective published papers as \textbf{AI-generated abstracts};
(3) We use the term \textbf{provided abstracts} to indicate those abstracts that we provide the authors to edit. Provided abstracts could be either original abstracts (in Human-noInfo or Human-withInfo treatments) or AI-generated abstracts (in AI-noInfo or AI-withInfo treatments). 
(4) We use the term \textbf{edited abstracts} to indicate the edited versions of the provided abstracts. To clarify, these edits have been made by the recruited authors. 
Our investigation includes two participant pools: Authors and Reviewers. We discuss the tasks, incentives, and recruitment procedures of each pool in the respective sections. 

\subsection{Authors}
\subsubsection{Task} 
% \newline
The main task of the authors is to make edits to the provided abstracts. Authors complete three tasks. In Task 1, we elicit authors' perceptions about their LLM use. The authors answer four questions. In the first question, we ask if authors use generative AI systems (e.g., ChatGPT) for any writing tasks. Participants choose between four options: they use generative AI to solely generate content, they use generative AI to only edit existing content, they use generative AI to both generate and edit content, or they do not use generative AI for any writing tasks at all. In the second question, we use an incentive-compatible method to elicit authors' second-order beliefs about the overall use of GPT in writing tasks. We ask what percentage of other authors they believe use GPT to generate new content or edit existing content. Authors are informed that if their selected percentage bracket matches the actual percentage (as calculated by the experimenter), they will receive a bonus of \$0.50. In the third and fourth questions, we ask authors to rate the performance of AI systems in generating and editing content, respectively.

\begin{figure}[!t]
    \centering
    \includegraphics[width=\linewidth]{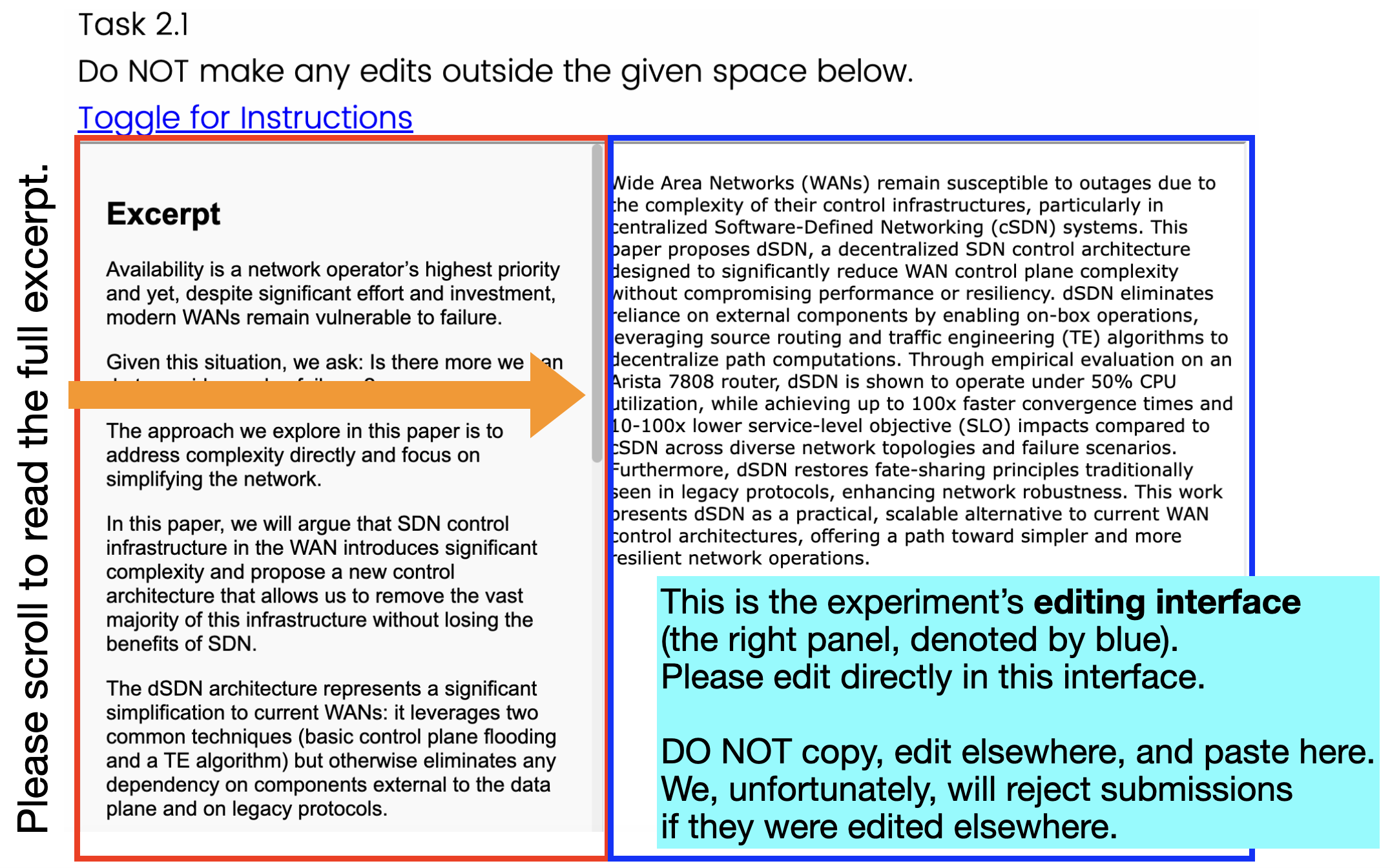}
    \caption{An example of the first detailed pictorial representation where experimenters show the authors the author panel and provide them explicit information that edits need to be made directly in the interface. We conducted several pilot studies where we found that the recruited authors were copy-pasting the provided abstract and making edits elsewhere. Providing these graphic instructions and additional screening questions helped mitigate this problem and enabled capturing keystroke-level edits for every abstract.}
    \label{fig:author_warnings1}
\end{figure}

\begin{figure}[!t]
    \centering
    \includegraphics[width=\linewidth]{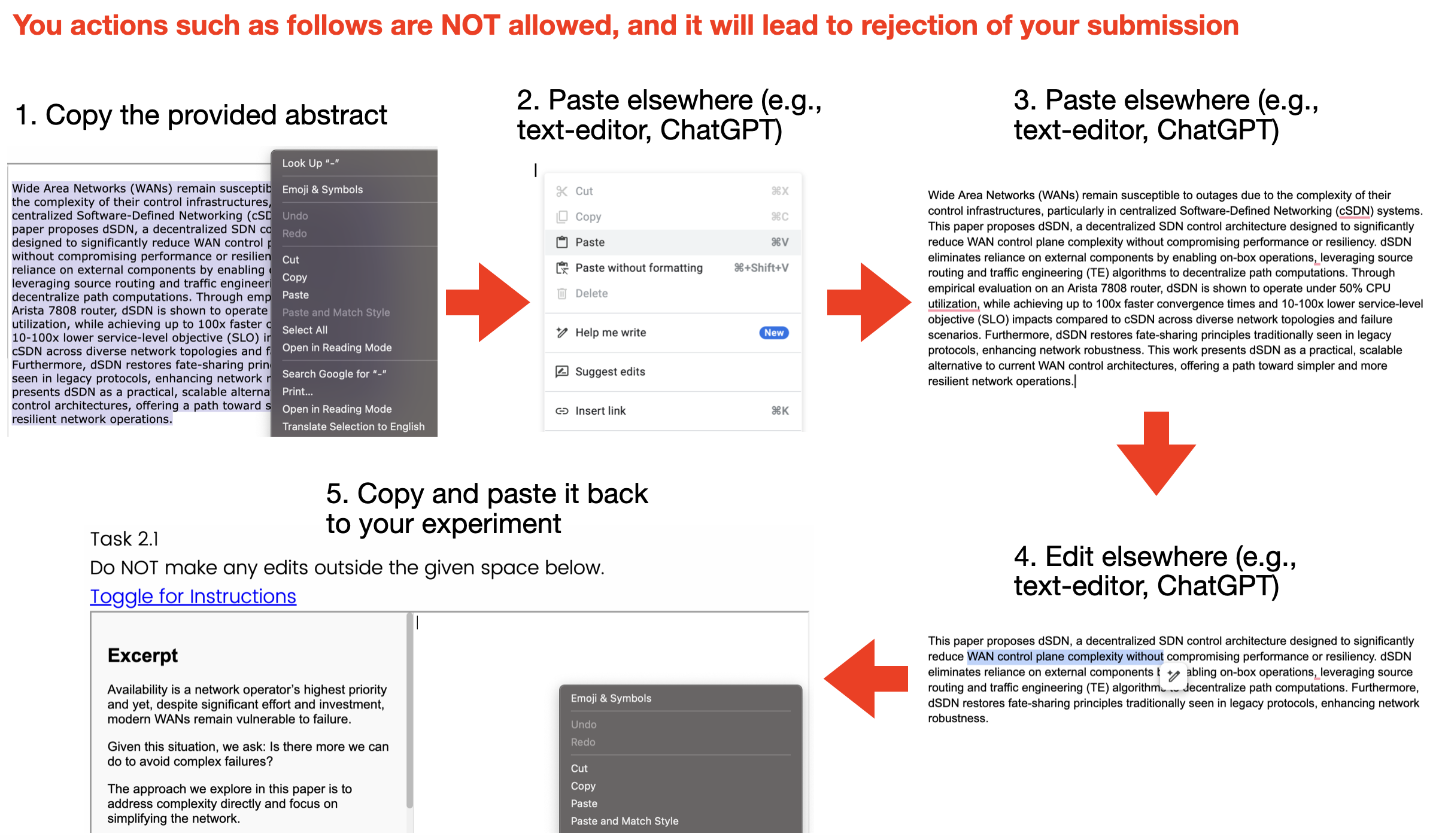}
    \caption{ Illustration of prohibited actions during the abstract editing task. Authors in the second schematic illustration are instructed not to copy the provided abstract into external tools (e.g., text authors, AI assistants), make edits outside the designated experimental interface, or paste modified content back into the system. We inform authors that such actions violate the experimental protocol and may lead to exclusion from the study. }
    \label{fig:author_warnings2}
\end{figure}

Task 2 entails the main editing task. Every author is presented with three independent research excerpts and their respective provided abstracts. The editing task appears to the authors as illustrated in Figure \ref{fig:authorinterface}. Figures \ref{fig:author_warnings1} and \ref{fig:author_warnings2} provide the detailed illustration provided to the authors to help them understand the instructions. Authors are instructed to first review the excerpts in the left panel of the screen. A scroll bar is provided to the authors to completely review the excerpt, which may extend beyond the scope of the current screen space. Simultaneously, the right panel entails the editing interface that shows the provided abstract. We explicitly instruct the authors to make all edits in the editing interface only. We disabled copy-pasting from the editing interface so that the participants cannot copy and paste to edit on a platform other than the one provided to them. Additionally, this will also limit authors from generating or making edits using existing AI models. The main goal of the authors is to receive \textbf{at least two Accept decisions} by the reviewers on their edited abstracts. Before starting with Task 2, the authors answer two screening questions (\autoref{fig:screening_questions}). These questions were designed to assess the authors’ understanding of task instructions. Correctly answering these questions was a prerequisite for moving forward with the main editing task. 

After the screening questions, the authors go through each excerpt-provided abstract combination and make relevant edits.
% \sk{should we mention here about the control and revealing information of AI vs human generated?} 
After the authors have made their edits, we elicit two confidence metrics from the authors in an incentive-compatible method. First, we ask the authors to evaluate the provided abstract and indicate how likely they believe it is to receive at least two Accept decisions by the reviewers. In the next question, authors are asked to evaluate the edited abstract, which now includes their own revisions, and indicate how likely they believe it is to receive at least two Accept decisions from reviewers. Responses for both questions are recorded on a slider scale ranging from 0 (not likely at all) to 50 (highly likely). An author edits \textbf{three} instances of provided abstracts and submits respective confidence questions for the three edited abstracts. We also ask authors to rate the quality of the provided abstracts (excellent to poor), describe their frequency of edits, mark the kind of edits authors have made (e.g., addition, deletion, correction, substitution, and reordering), and finally rate the readability of the provided abstracts, using a slider scale where 0 indicates no readability and 100 indicates complete readability.

Authors proceed to Task 3 after they have made their edits and submitted the final versions of edited abstracts. Task 3 involves completing an exit questionnaire that gathers participants' demographic information (such as gender, education, current profession, age), their first and second-order beliefs about using AI to write abstracts or scientific reports, familiarity AI-generated text versus human-written text, experience with AI tools in daily life, trust in AI tools, and frequency of AI tool usage for writing tasks.

% \vfill

\subsubsection{Incentives}
% \newline
% \sk{need to mention median/average time it took to finish the task.}
Authors earn a fixed participation fee of \$15 for participating in this study and completing all the tasks. Note, authors receive this payment irrespective of their abstracts being accepted by the reviewers or not. Using the methodology of economics experiments, we additionally incentivize the participants to receive a bonus pay of \$15 if a randomly selected abstract out of the three edited abstracts receives at least two Accept decisions from the reviewers. Authors also receive a bonus for their answers in the confidence questions, respectively, according to their submitted confidence levels. E.g., if an author marks 20 on the slider for the edited abstract and it receives at least two Accept decisions from the reviewers, the authors earn a bonus of \$0.20. An author can receive a maximum of \$0.50 from each of the confidence questions. 

To encourage genuine effort and careful editing, our incentive structure followed conventions from experimental economics, where performance-contingent rewards are known to elicit higher attention and task quality \citep{smith1982microeconomic,camerer1999effects}. The fixed participation fee ensured baseline engagement, while performance-based bonuses motivated authors to produce edits that improved clarity and persuasiveness rather than completing the task superficially \citep{abeler2011reference}. 
%A small confidence-based bonus further encouraged accurate self-assessment, helping us interpret reported confidence as a meaningful behavioral signal \citep{offerman2009truth}. 
This helps us maintain data quality and internal validity while aligning with best practices in incentive-aligned experimental and crowd-sourcing studies \citep{horton2010labor,mason2009financial}. We do not design confidence questions to be an incentive-compatible belief elicitation. Instead, our bonus serves as a small, performance-contingent incentive to increase effort. Since the acceptance is probabilistic, overstating confidence without producing a high-quality edit does not increase expected payoffs.

In total, an author could earn a guaranteed amount of \$15, with the possibility of an additional maximum bonus reward of \$16.50 (i.e., a maximum total of \$31.50). The authors were paid after the completion of the study. After completing all three tasks, the participation fee was immediately transferred to the authors. For the bonus, the computer randomly selected one edited abstract out of the three that the authors submitted for the hypothetical conference set-up, and the authors were informed that they would be paid their bonus based on the reviewer's decisions for the selected abstract. The average and median time taken for this task were both 1 hour and 8 minutes; our payments justify the stipulated hourly rates on Prolific.

\subsubsection{Procedures and Author Recruitment}
We recruit 300 authors via the online crowd-sourcing platform \href{https://www.prolific.com/}{Prolific}.\footnote{This research has been IRB approved by the Office of Responsible Research Practices at Ohio State University under study number 2024E1034. We also pre-registered for our trial at AEA RCT Registry: \href{https://doi.org/10.1257/rct.16740-1.0}{https://doi.org/10.1257/rct.16740-1.0}} We use the \href{https://www.qualtrics.com/}{Qualtrics} interface to set up our study, which was then linked to Prolific for every author. Participation in the study was voluntary. At the onset of the experiment, each author was shown information about the study, followed by an informed consent form.

For our analysis, we exclude 3 authors owing to malformed responses, resulting in 297 authors. Limiting our study to only the CS domain limits us to restricting authors based on their total number of study approvals on Prolific. The median total approval number for our authors is 336.0 approvals, with a minimum of 0 and a maximum of 5560 approvals. 
% We only include authors who were located in the United States. 
88.51\% of authors have their registered first language as English. On average, the authors' ages (self-reported) range from 25 to 40, and 42.42\% of the authors are female (self-reported).

Since all our source material for creating the abstracts is strictly limited to the CS domain, we restrict authors to belonging to the CS domain by prescreening them based on their reported \textit{Subjects} (proxy for domain of study), as recorded on Prolific. It is a general norm in the highly competitive field of CS that students are engaged in scientific writing since their undergraduate years. To optimize for authors to have some experience with formal writing in CS, we also filter authors by their \textit{Highest education level completed}. Authors include individuals who have an undergraduate degree (BA/BSc/other), a graduate degree (MA/MSc/MPhil/other), and/or a doctorate (PhD/other). 80.00\% of authors are currently maintaining their student status, with 55.56\% of authors having at least an undergraduate degree with a major/minor in CS.

\subsection{Editing Interface}
\label{sec:editinginterface}

Our primary evaluation and analyses are based on recording edits made by the authors. To store these edits, we explored multiple popular text editing tools that provide versioning history, such as Google Docs and Microsoft Word. 
% There were different options for documenting text edit history. The most widely used author is Google Docs. 
% While they is a reliable platform with version history provided by their API, 
Having faced several challenges with these tools, we were encouraged to build our own interface. For example, these platforms do not store every keystroke. Instead, they record versions at a fixed time interval, which cannot be modified by the user. This results in us losing many edits that were valuable for our analysis.  %the frequency of version history storage for most platforms is   is unclear and the edit history is saved by bulk of edits, not by keystroke. 
Furthermore, these platforms do not allow finer-grained control over how the edits can be made. For example, to prevent authors from using AI-assistance to make edits, we needed to disable copying, which none of the existing platforms allow.
% As we need to conduct in depth analysis which requires a fine-grained edit log, Google Docs was not adopted in our study. 
We also explored Revision History, %our subsequent target application, is 
a Google Docs Extension that offers full document composition record at character-level. This tool offers the magnitude of editing details we sought. However, it operates entirely within Google Docs, is handled online, and therefore does not support exporting the edit history. %Subsequently, we intended to separately export the same edit history of Google Docs that Revision History were accessing and integrate their parsing logic when interpreting the history. Google Docs does maintain fine-grained edit history, but exporting it has to be conducted through accessing the data of the document webpage as it is not provided by the API. 
% However, 
Google also prohibits automated web scraping of their data, and thus, this pipeline was not reproducible for mass edit log collection. 

% Under these limitations, our next target was Firepad, 
To build our own interface, we used Firepad,
an open-source text editing tool that stores detailed edit history on Firebase Realtime Database and uses CodeMirror as the underlying text editor. While Firepad is no longer being hosted globally, we used its open-source code\footnote{\url{https://github.com/FirebaseExtended/firepad}} to host it ourselves through Firebase Web Hosting. Throughout this study, we used Firepad to record the time, the index, and the edited content per keystroke. We implemented our interface using Firepad 1.5.11, Firebase 5.5.4, and CodeMirror 5.17.0. We plan to open-source this interface for ease of reproduction and use by other researchers. %versions for implementation. 
For each author and abstract pair, we create unique Firepad pages. The combination and the index of each pair are used as URL parameters to dynamically populate the authors. Each Firepad page is structured as research findings or excerpts on the left and provided abstracts on the right (\autoref{fig:authorinterface}). To allow authors to refer to the provided abstract when necessary, even though they made changes to the provided abstract, we include the provided abstract at the bottom of the excerpts section. Excerpts, in HTML, are non-editable and embedded in the page. Provided abstracts are pre-loaded in the editing box. We embed the Firepad links into a Qualtrics\footnote{\url{https://www.qualtrics.com/}} survey form.

\subsection{Reviewers}

\subsubsection{Task}

Reviewers complete three tasks. Task 1 for reviewers is identical to the Task 1 completed by the authors. We elicit reviewers’ perceptions about their LLM use.
Task 2 entails reviewing abstracts. Each reviewer is presented with a set of 20 abstract pairs---the original abstract that accompanied a selected research paper and the respective edited abstract by an author.\footnote{Showing the original abstract in comparison to the edited versions, eliminates any potential bias in reviewers’ abstract acceptance that may arise from the subjective content of the research.} We inform the reviewers that the edited abstracts encapsulate the same research idea, with the same research methodology and results as the original abstracts. However, it may or may not be expositionally different from the original abstract. Reviewers need to review the edited abstract in reference to the original abstract and answer the following question: \textit{Does the edited abstract provide adequate justice to the research idea presented in the original abstract?} The reviewers answer using a slider that ranges from 0 (no justice, worse than the original abstract) to 100 (better than the original abstract), with the default set at 50 (i.e., both abstracts can be considered equally good and one is not better than the other). We also elicit the absolute confidence of the reviewer in each of their 20 decisions. Additionally, to get more insights into the decision-making process of the reviewers, we provide a text box where the reviewers can address this in free-form text. % in a few lines. 
Reviewers proceed to Task 3 after completing Task 2. Task 3 for reviewers, identical to Task 3 completed by the authors, comprises completing an exit questionnaire.
We ask two additional questions about the publication record of the reviewers and their prior knowledge of the topics discussed in the abstracts. Note, we do not disclose the source of the edited abstract to the reviewers.\footnote{In our setup, we aim to isolate the reviewers' perception of the quality of edits in the edited abstracts compared to the original. We do not want to confound acceptance decisions with reviewers’ inherent beliefs about AI-generated scientific abstracts.}

\subsubsection{Incentives}

% \sk{need to mention median/average time it took to finish the task.} 
Reviewers earn a fixed participation fee of \$25 for participating in this study and completing all the tasks, irrespective of their decisions in the tasks. Leaning back onto the methodology of incentivized experiments, to increase data quality, we provide an additional bonus of \$8 to the reviewers whose evaluations align closely with those of their peers. One of the twenty abstracts is randomly selected, and two independent reviewers, alongside the reviewer-in-question, assess the same abstract pair using the slider scale described in Task 2. We then compute the median of the three ratings, and if a reviewer’s score falls within 10 points above or below this median, they receive the \$8 bonus. This peer-alignment incentive helps ensure careful, calibrated judgments and reduces random or inattentive responding. Unlike Task 1 incentive for authors, reviewers earn a bonus of \$2 from Task 1, task remaining identical. The first quartile of time taken for the reviewing task is also 1 hour and 8 minutes. However, the reviewers were not required to complete the task during one continuous sitting, and therefore, many took more than a day. Payments to reviewers were made through registered Amazon gift cards. 
% (this is for 95 reviewers from recent sets)}

\subsubsection{Procedures and Reviewer Recruitment}

We have a total of 159 reviewers. We recruited the reviewers from the current student population at universities who are currently pursuing their graduate and doctoral studies to review 891 abstracts. 174 students signed up for reviewing using the distributed recruitment form. However, only 118 finally took part and completed the study\footnote{The first set of 60 reviewers reviewed nine abstracts and received a flat payment of \$15, with a bonus of \$0.50 from Task 1. To maintain a high-quality of the reviewers, we maintained a restricted pool, which required us to increase per-reviewer assignment as well as hourly payment.}. 41 remaining reviewers were recruited on Prolific. On Prolific, we maintain the same restrictions for recruiting reviewers as authors; however, the reviewers are now restricted to having completed the highest education level of a graduate degree (MA/MSc/MPhil/other) and/or a doctorate (PhD/other). All reviewers were shown information about the study prior to starting the study, followed by an informed consent form.

32.70\% of the reviewers are female (self-reported), 73.59\% of reviewers hold at least a master's degree, 89.94\% of reviewers either have a publication or are in progress, with 66.04\% having prior knowledge regarding the topics discussed in the abstracts. We have 112 reviewers who provided a Google Scholar or Semantic Scholar link. The median number of publications is 6.5, with a minimum of 1 and a maximum of 61. The median number of citations is 32.5, with a minimum of 0 and a maximum of 4974. For reviewers recruited on Prolific, the median total approval number is 367, with a minimum of 2 and a maximum of 6795 approvals.

\begin{figure*}[t!]
  \centering
  % ---------- first panel ----------
  \begin{subfigure}{0.32\linewidth}
    \includegraphics[width=\linewidth]{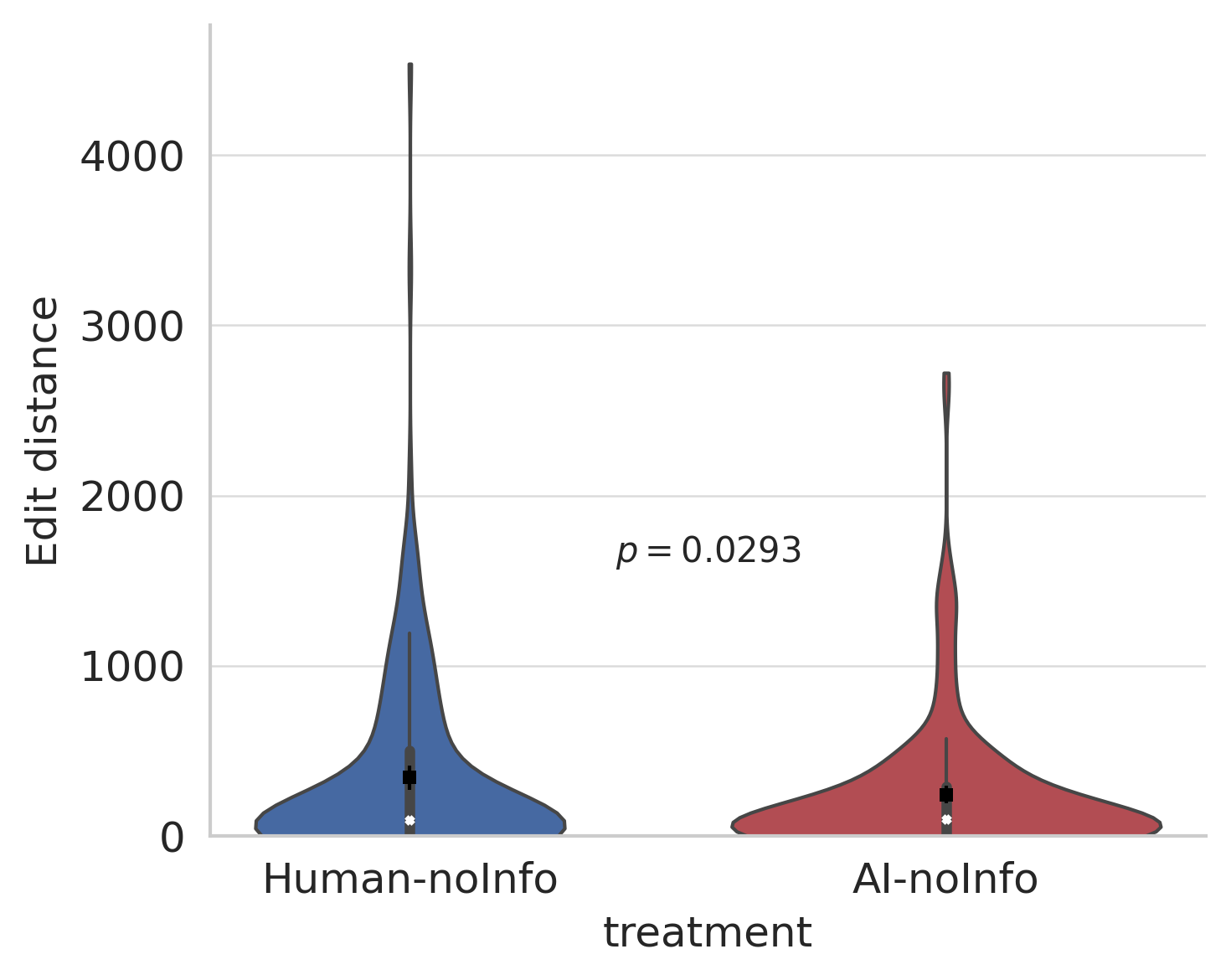}  % or .png/.jpg
    % \caption{Panel A}
    % \label{fig:panelA}
  \end{subfigure}
  \hfill
  % ---------- second panel ----------
  \begin{subfigure}{0.32\linewidth}
    \includegraphics[width=\linewidth]{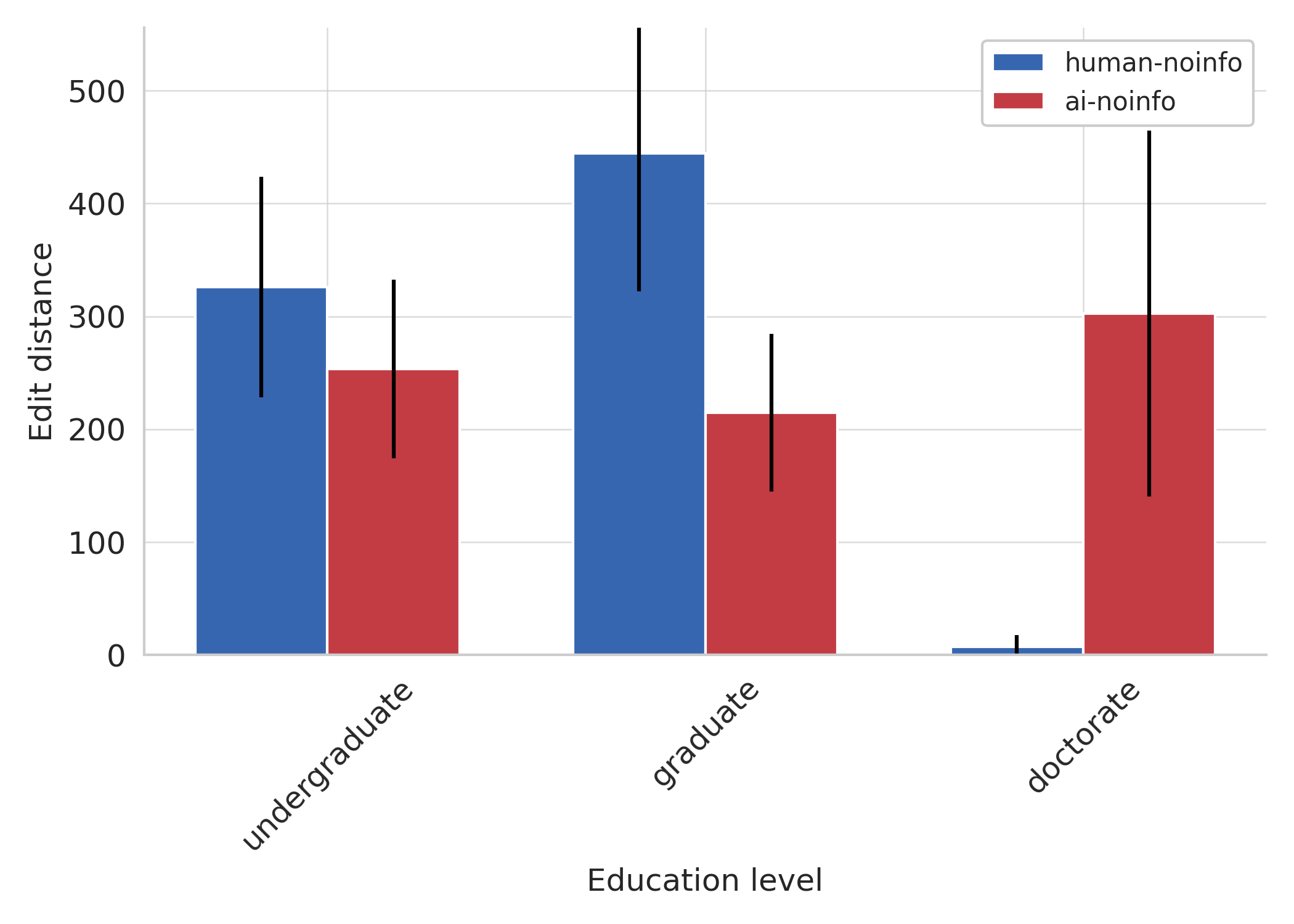}
    % \caption{Panel B}
    % \label{fig:panelB}
  \end{subfigure}
  \hfill
  % ---------- third panel ----------
  \begin{subfigure}{0.32\linewidth}
    \includegraphics[width=\linewidth]{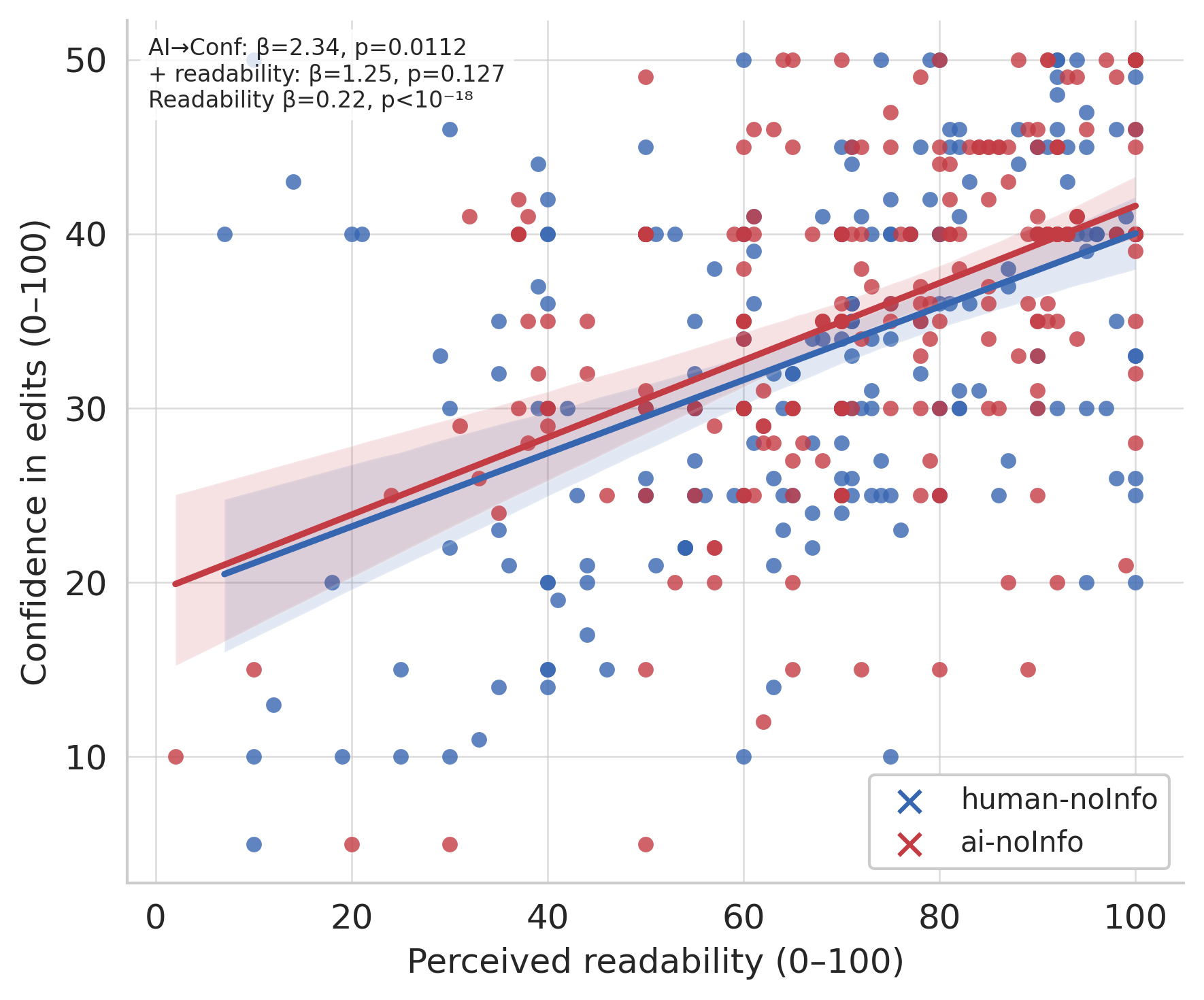}
    % \caption{Panel C}
    % \label{fig:panelC}
  \end{subfigure}
  % ---------- overall figure caption ----------
  \caption{Patterns of editing effort in Study 1 (N = 495 abstracts). \textbf{\textit{Left}:} Distribution of Levenshtein edit distance in two conditions: Human-noInfo (blue) and AI-noInfo (red). Black squares mark the sample means ± 95 \% CI; the Welch unequal-variance test indicates a statistically reliable reduction in edits for AI abstracts (p = 0.0293). \textbf{\textit{Middle}:} Mean edit distance by the authors' highest education level. Authors with undergraduate or graduate degrees make markedly larger edits to human text than to AI text, whereas doctorate-level editors show the opposite. \textbf{\textit{Right}:} Scatter-plot of abstract’s perceived readability (0–100) vs the author’s confidence in their edits (0–100). The positive slopes show that authors feel more confident when an abstract reads more smoothly; readability, not AI, as the source explains authors’ higher confidence in the AI-noInfo condition.}
  \label{fig:study1}
\end{figure*}

\subsubsection{Reviewer Assignment}

We adopted an automated pipeline to assign abstracts to recruited reviewers by mimicking the in-practice reviewer assignment mechanism in OpenReview, adopted by several CS conferences \cite{stelmakh2023gold}. In addition to topical fit, we also try to ensure that no reviewer can review more than one version of an edited abstract originating from the same original abstract to avoid contamination. First, we collect each reviewer’s scholarly profile using the Semantic Scholar APIs. Then we compute the semantic similarity between reviewer publications and base original abstracts using GPT-4o. We account for the reviewer's citation count to weigh the similarity between an individual paper and submitted abstracts, so that we assign related abstracts closer to the reviewer's most cited works. Finally, we solve the reviewer assignment as a constrained optimization problem by a minimum-cost flow, with constraints that each submitted abstract gets exactly three reviewers and each reviewer preferably does not review edit versions of the same original abstracts.

\subsection{Computed Variables}
We additionally compute the following variables for our analysis.
For the Author pool, we collect every character-level edit that the author makes to the original abstract provided to them. From here, we compile
% three metrics: (1) \textbf{Character-level edits} consisting of the number of character deletions, insertions (or both) made to the original abstract; (2) \textbf{Word-level edits} including the number of words deleted from, inserted in (or both) the original draft; (3) 
\textbf{edit distance score} measures the character-level Levenshtein distance\footnote{\url{https://en.wikipedia.org/wiki/Levenshtein\_distance}} between the original and edited abstract. This is the main variable of interest. For the reviewers, for each abstract, we collect a score between 0-100 indicating if the edited abstract does justice to the original abstract. If the score is $> 50$, then we binarize that decision to be "Accept" (else "Reject") since the original abstract is from a published work. For each edited abstract, three independent "Accept"/"Reject" decisions are collected, and we compute a \textbf{final decision} ("Accept"/"Reject") using the majority voting rule. The binarized final decision is the same variable that we use to determine the author bonus. 
% utilize this decision, at an abstract level, to finalize the final outcome of the abstract, i.e., if at least two reviewers independently accept an edited abstract, we say that the edited abstract has been accepted for presentation in the hypothetical conference; otherwise, the edited abstract has been rejected.\sk{mention the threshold for accept}

    % % \item 
    % Confidence measures: We also capture how confident the Authors and Reviewers are in their ability to edit and review the abstracts.
% \end{enumerate}

% \section{Study A: Authors}\label{sec:authorsdesign}

% \subsection{Study B: Reviewers}\label{sec:reviewerdesign}

% \newpage
% \noindent
% \textcolor{red}{\textbf{Bodhi adding new sections from here}}

\section{Study A: Human-written vs AI-generated Abstracts}
\label{sec:studya}

Study A focuses on exploring how authors evaluate and make edits to scientific writing, irrespective of the source information of the script. In a between-subject experiment, we compare two treatments, \textbf{Human-noInfo} and \textbf{AI-noInfo}. Authors are uniformly randomly assigned to either of the two treatment groups. Authors in Human-noInfo treatment edit the original abstracts from the selected published research paper, whereas the authors in the AI-noInfo treatment edit the pre-populated AI-generated abstracts of the selected published research paper. Authors do not have any information about the source of the abstracts and are only provided the excerpts and the abstracts to make their edits. This comparison allows us to observe the natural editing behavior of authors free from disclosure-driven bias. By removing any source cues, we observe how individuals naturally engage with a piece of scientific writing, how they judge its clarity, coherence, and overall quality without being influenced by who (or what) produced it.

% We induce a $2\times2$ between-subjects experiment design with four treatments. One dimension of the design includes two information conditions: \textbf{No Information} and \textbf{With Information}, representing the absence and presence of source information of the abstracts. Within each information condition, we vary the source: \textbf{Human-Written} and \textbf{AI-Generated}. Human-written indicates that the abstracts are the original abstracts from the selected research papers written by experts in the subfield. AI-generated implies that the abstracts have been generated using an LLM.
% The treatments are thus labeled as: Human-Written No Information (Human-NoInfo), Human-Written With Information (Human-WithInfo), AI-Generated No Information (AI-NoInfo), and AI-Generated With Information (AI-WithInfo). A total of 60 participants (authors) took part in the study as members of the Author Pool, with 15 authors in each treatment. Each author was randomly assigned to only one treatment.

% In this study, we want to understand if 

% \subsection{Conditions}
% We have two conditions.

% \paragraph{Human abstracts}

% \paragraph{AI-generated abstracts}
% We use gpt-4o...

% \paragraph{Hypotheses}

% \begin{quote}
% 1. Authors edit the AI-generated abstract more.\\
% 2. Less-edited AI-generated abstracts will be less accepted.
% \end{quote}

\subsection{Results}

% \textbf{Overall edit:} The analysis reveals a consistent difference in editing intensity across abstract sources. Participants made substantially larger edits to human-written abstracts ($M = 279.34$, $SD = 326.96$) than to AI-generated ones ($M = 227.91$, $SD = 270.74$). A Welch’s \textit{t}-test confirms that this difference is statistically significant, $t(\approx 880) = 2.56$, $p = 0.011$, indicating that human-origin texts prompted more extensive revisions. The distribution overlap, however, suggests that participants’ editing effort varied widely across both types, reflecting heterogeneous engagement styles rather than a uniform aversion or enthusiasm toward AI-generated prose.

\subsubsection{Overall edit} When authors were unaware of whether an abstract was written by a human or an AI system, we found systematic yet heterogeneous differences in editing patterns. We use ordinary least squares (OLS) regression with heteroskedasticity-consistent standard errors to estimate that the authors made significantly smaller edits to AI-generated abstracts compared to human-written ones, on average, about 63 characters fewer (see \autoref{fig:study1}, Left). When the source of the text is unknown, authors perceived AI-written text as closer to a submittable form with lower intent to modify it.

\subsubsection{Heterogeneity through education} To account for differences across individual authors, we consider demographic characteristics such as gender and education, along with their interactions with the treatments. Authors with doctoral degrees made substantially more edits to AI-generated abstracts ($p = 0.012$; joint Wald test), reversing the average effect. On the contrary, those with undergraduate or master’s degrees made smaller changes (see \autoref{fig:study1}, Middle). This pattern hints that academic (writing) experience promotes the ability to identify idiosyncrasies of AI-generated text. In contrast, less experienced authors interpret the same fluency as high quality, leading to fewer definitive edits. 

To probe further, we estimate a mixed-effects model for both author and abstract, capturing unobserved heterogeneity in editing intensity. The model finds negligible between-abstract variance (intra-class correlation $= 0$), suggesting that most variation comes from idiosyncratic styles rather than intrinsic differences in abstract quality. In an auxiliary model using \emph{perceived readability} as a moderator, higher perceived readability strongly predicts lighter editing ($p < 0.001$). This does not significantly interact with the AI treatment. We find that the baseline content features of AI-written text shape how authors edit, and that this behavior depends critically on authors’ writing expertise rather than the inherent readability of the content.

Furthermore, authors' prior familiarity with AI tools, trust in AI, and expectations about AI performance show little influence on how extensively authors edit the abstracts. None of the interaction terms between experience with AI and AI-treatment are statistically significant, suggesting that even frequent AI users or authors who highly trust AI do not edit AI-generated text differently from those less experienced with AI. The main treatment effect remains robust and negative ($\beta = -64.9$, $p = 0.038$). Authors make substantially fewer edits to AI-generated abstracts, but the observed differences stem primarily from intrinsic properties of the text (e.g., readability), and not from authors’ prior beliefs on AI.

\subsubsection{Ability to distinguish AI-generated text partly explains edits}
We next examine whether authors’ self-reported ability to distinguish AI- from human-written text shapes their editing behavior. We focus exclusively on the \textit{AI-noInfo} condition, since we elicit authors' ability to distinguish AI vs human-written text in the AI treatment only. We compare the edits of authors who reported being able to distinguish AI from human text (“Yes/Quite often”) with those who reported being unable or unsure (“Rarely/Never”). Authors who claim they could distinguish AI-generated writing made larger edits on average ($M = 257.03$, $SD = 316.38$, $N = 198$) than those who could not ($M = 157.56$, $SD = 204.57$, $N = 27$), the difference being statistically significant ($p = 0.033$; Welch’s \textit{t}-test). To account for potential confounds, we regress edit distance on self-reported ability to distinguish AI text along with original confidence, perceived readability, and demographic controls (gender and education). The effect of AI distinguishability remains positive but becomes statistically insignificant ($\beta = 97.15$, $p = 0.124$, $N = 135$), which indicates the initial difference is partly explained by correlated covariates. Authors with higher pre-edit confidence tend to make smaller edits ($\beta = -6.20$, $p = 0.074$), while perceived readability does not significantly predict edit distance ($p = 0.990$). This implies that authors who believe they can identify the difference between AI and human writing tend to edit AI-generated text more extensively, but this effect is not independently robust once we control for individual confidence and perceived readability.

\subsubsection{Perceived readability mediates confidence}
Authors express consistently higher confidence in their edits on AI-generated abstracts ($\beta = 2.34$, $p = 0.011$ from an OLS) compared to human-generated ones when they remain unaware of the source of the text. However, this effect reduces significantly once authors’ \textit{perceived readability} is included in the model. When we add perceived readability as a covariate, the coefficient on the AI indicator drops by nearly half and becomes statistically insignificant ($\beta = 1.25$, $p = 0.123$), while perceived readability itself emerges as a strong positive predictor of confidence ($\beta = 0.22$, $p < 10^{-18}$), as shown in \autoref{fig:study1}, Right. Authors rate AI-generated abstracts as more readable on average ($M = 73.01$, $SD = 19.91$) than those written by humans ($M = 67.94$, $SD = 22.80$), and perceived readability is moderately correlated with confidence ($r = 0.48$). Together, these patterns indicate that readability perceptions substantially moderate and explain the higher confidence associated with AI-generated text. Once we account for readability, the direct effect of the AI source largely disappears. 
% Authors’ initial confidence is driven not by implicit trust in AI, but by the impression that AI text is smoother, more coherent, and easier to process.
Even after including demographic controls (gender and education), perceived readability remains a strong and stable predictor of confidence ($\beta = 0.19$, $p < 10^{-9}$), whereas the AI effect remains modest and only marginally significant ($\beta = 2.24$, $p = 0.024$). The perception of how readable the abstract is mediates how the author gains confidence in their edits.

\subsubsection{Reviewer decisions}
Finally, we model final acceptance decisions for Human-noInfo and AI-noInfo conditions. We find no difference in acceptance probability (AI: $\beta=-0.198$, $p=0.346$), indicating authors are able to bring the abstracts to an acceptable quality when starting from either AI- or human-authored versions when the source is undisclosed.

\section{Study B: Source Information Disclosure}
\label{sec:studyb}

The source of text, in recent times, could be human-written or AI-generated. While detection is often challenging, disclosing the source of a scientific text could be fundamental in shaping individual evaluation and interaction patterns with the text. Individuals have often reported reading AI-generated text more critically, anticipating errors, and focusing on the factual accuracy to encounter model hallucinations. In contrast, knowing that a text is human-produced can either invoke trust, or a reluctance to make additional edits. Individuals often feel a stronger sense of oversight when working with AI, or greater deference when editing human work. \citep{buccinca2021trust, li2024does} provides evidence that transparent source information is crucial to altering calibration, confidence, and task performance. 

By varying source disclosure, we can examine how people edit in light of source disclosure and why, whether their actions stem from intrinsic perceptions of quality or from social and cognitive biases linked to authorship cues. We have two disclosure treatments: Human-withInfo and AI-withInfo. In both treatment conditions, we inform authors of the source of the provided abstract. As outlined earlier, the provided abstract could either be the original abstract or the AI-generated abstract. 

% \subsection{Conditions}
% \begin{itemize}
%     \item Human-noInfo vs Human-withInfo
%     \item AI-noInfo vs AI-withInfo
%     \item Human-withInfo vs AI-withInfo
% \end{itemize}

% \paragraph{Hypotheses}
% \begin{quote}
%     \item \textcolor{red}{TODO}
% \end{quote}

\begin{figure*}[t!]
  \centering
  \begin{subfigure}[t]{0.32\linewidth}
    \includegraphics[width=\linewidth]{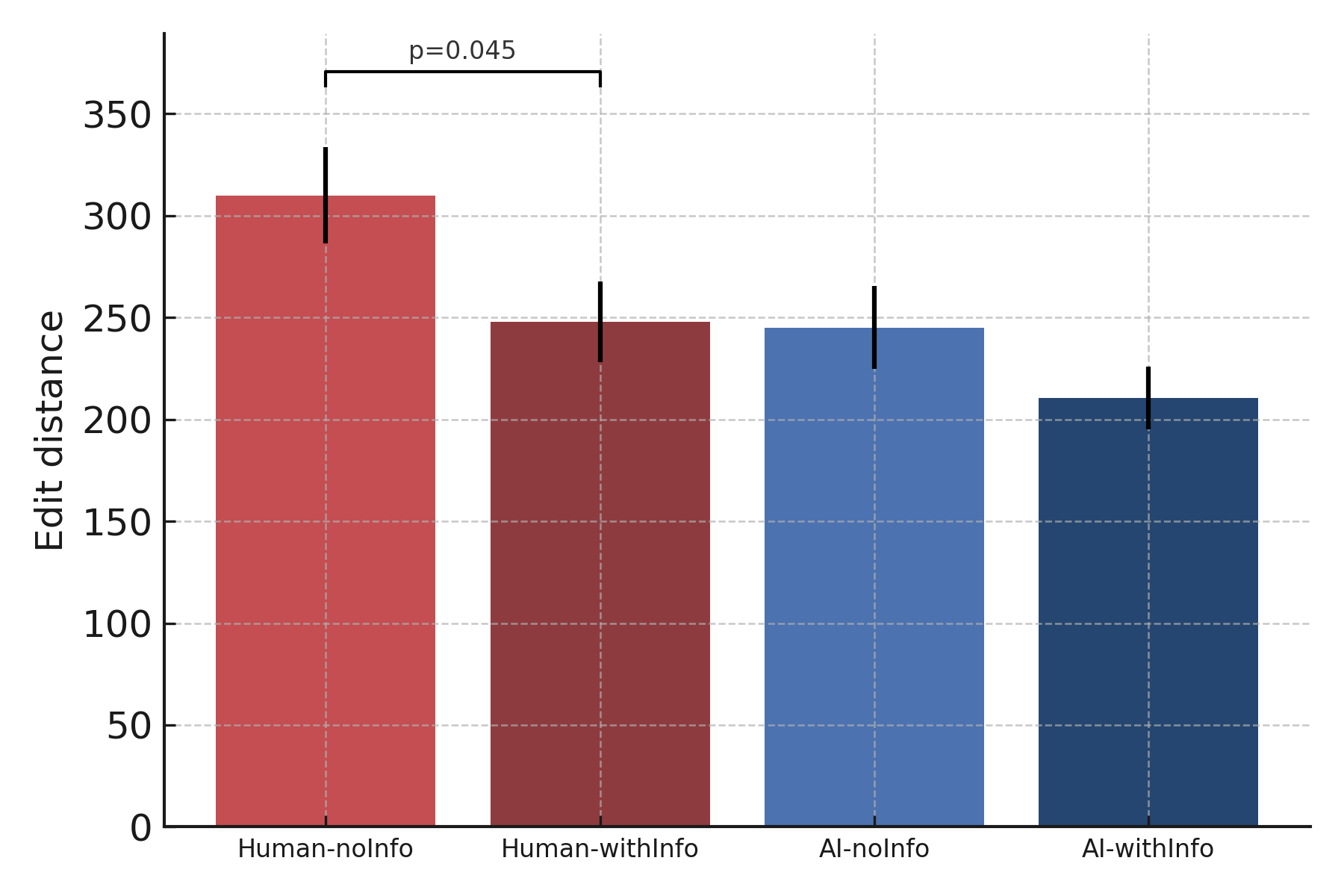}
    % \caption{Overall edit distance}
    % \label{fig:editdistance}
  \end{subfigure}
  \hfill
  \begin{subfigure}[t]{0.32\linewidth}
    \includegraphics[width=\linewidth]{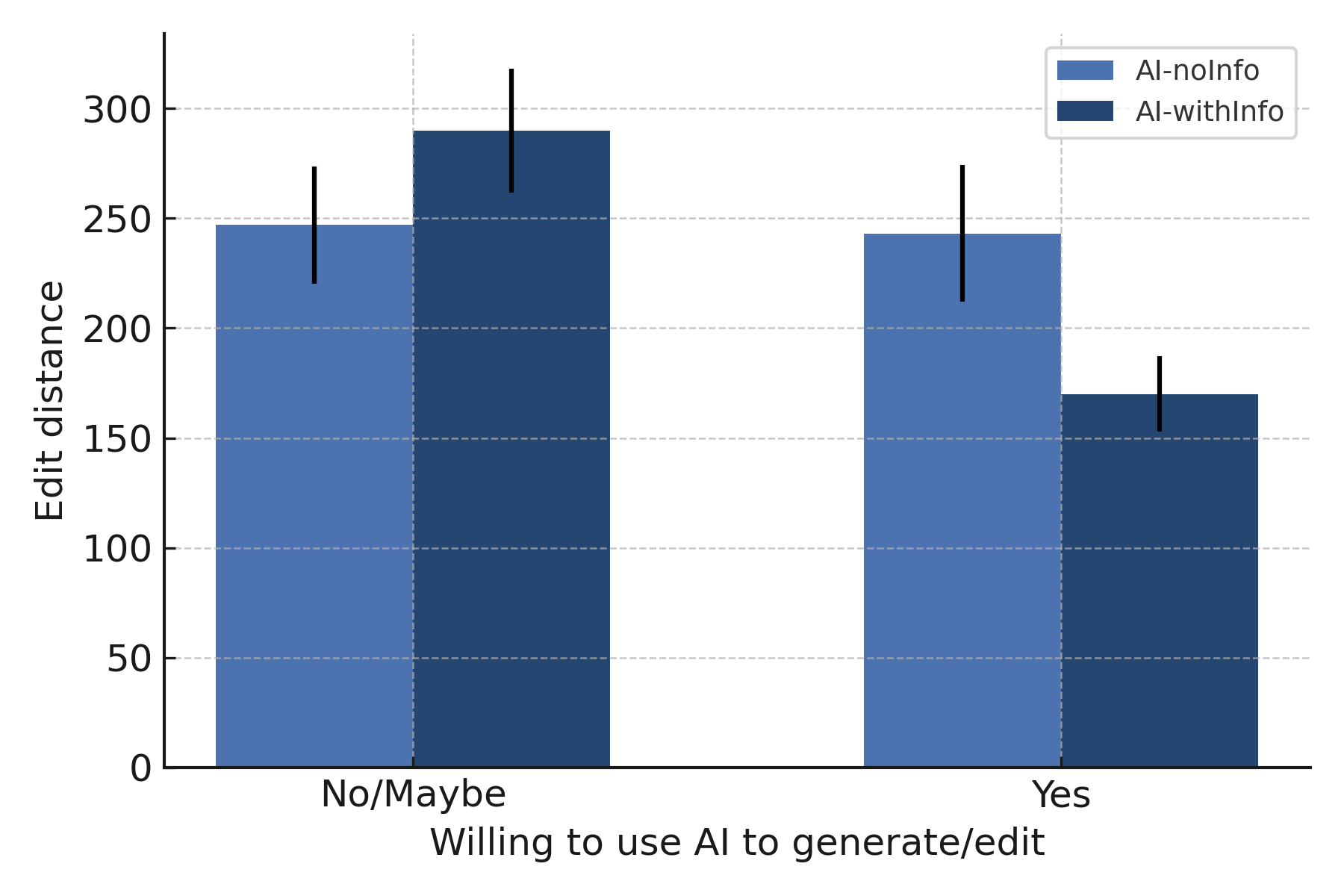}
    % \caption{Heterogeneity by AI willingness}
    % \label{fig:heterogeneity}
  \end{subfigure}
  \hfill
  \begin{subfigure}[t]{0.32\linewidth}
    \includegraphics[width=\linewidth]{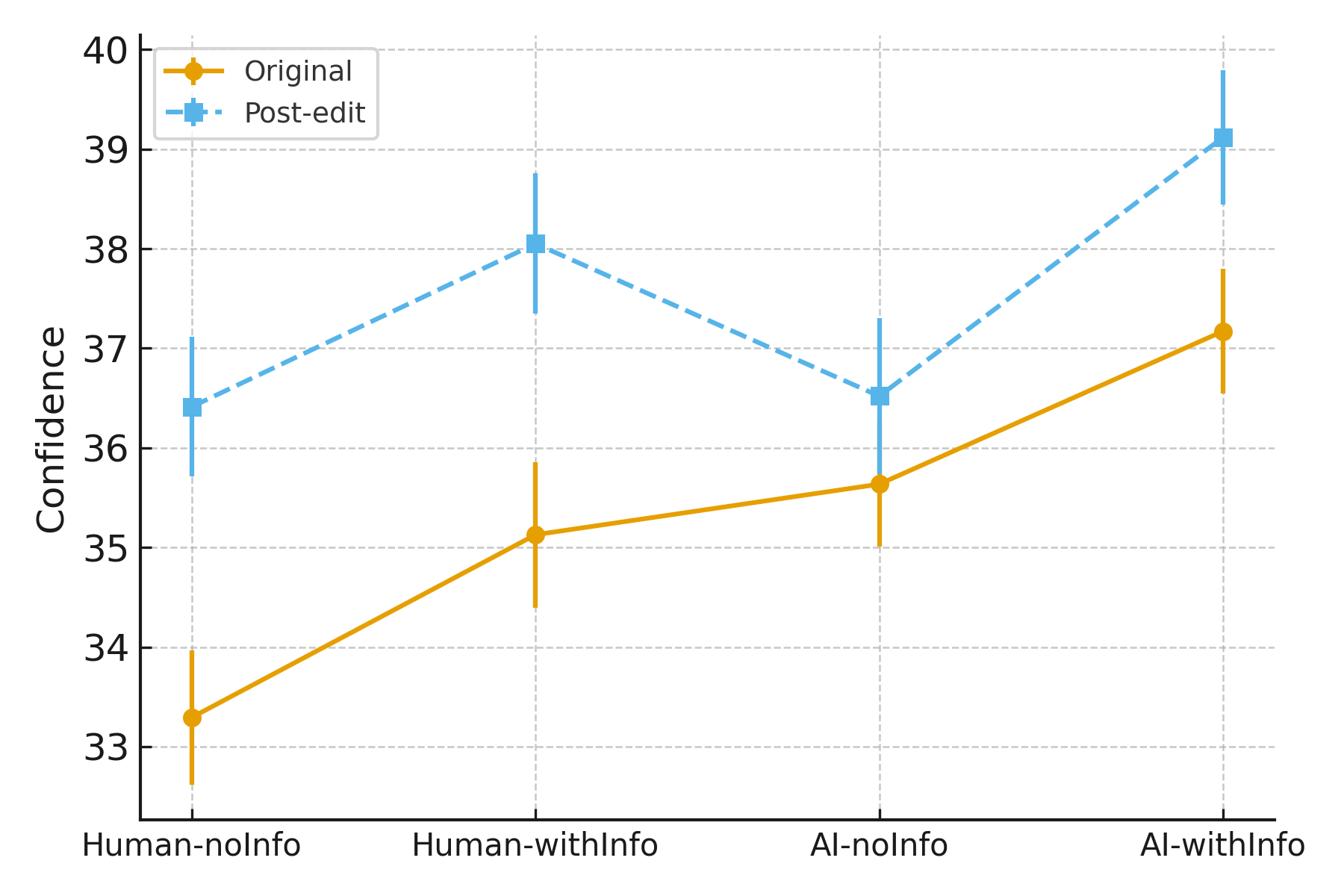}
    % \caption{Confidence before/after editing}
    % \label{fig:confidence}
  \end{subfigure}
  \caption{Treatment effects on editing behavior, heterogeneity, and confidence in Study 2 (N = 495 abstracts). \textbf{\textit{Left}:} Mean character-level edit distance (± SE) for each treatment condition. Disclosing human authorship (Human-withInfo) significantly reduced edits relative to the no-disclosure baseline (Human-noInfo; $p = .045$), whereas the analogous AI contrast was smaller and nonsignificant. \textbf{\textit{Middle}:} For AI-generated abstracts, the disclosure effect depends on whether authors say they would personally use generative AI. Among those not willing to use AI, disclosure had little impact; among those willing to use AI, edit intensity fell markedly in the AI-withInfo condition, highlighting that individual AI adoption attitudes shape editing effort. \textbf{\textit{Right}:} Disclosure of source as AI slightly lowers self-reported pre-edit confidence in acceptance but does not alter the post-edit confidence gains observed across all conditions.}
  \label{fig:study2results}
\end{figure*}

\subsection{Results}

\subsubsection{Overall edit} 
To examine how disclosure of source information influences the authors’ overall edit intensity, we compare the character-level edit distance pairwise across the final four treatments (Human-noInfo, Human-withInfo, AI-noInfo, and AI-withInfo).
% We applied Welch’s unequal-variance \textit{t}-tests for each pairwise comparison, reporting 95\% confidence intervals and Hedges’ \textit{g} as the standardized effect size. 
Authors make substantially larger edits when they were not told that the abstract was human-written ($M = 309.98$, $SD = 353.40$, $N = 225$). In contrast, than when source information is disclosed ($M = 247.87$, $SD = 294.86$, $N = 219$), there emerges a statistically significant difference, $t(431.96) = 2.01$, $p = .045$ (see \autoref{fig:study2results}, Left).  
Disclosure appears to dampen authors’ inclination to edit human-authored text. This proposes a restraint based on self-reputation once authorship is made known. A similar but nonsignificant pattern emerged for AI-generated abstracts. Edits were moderately higher ($p = .176$) in the no-information condition ($M = 245.09$, $SD = 306.50$, $N = 225$) than when AI origin was disclosed ($M = 210.50$, $SD = 228.27$, $N = 222$).  
As disclosure reduces the total edits, the smaller and nonsignificant effect suggests \textit{less sensitivity to provenance cues} when participants knew the text was AI-generated.  
Comparing the two conditions where the source information was disclosed always, authors tend to make larger edits to human-authored original abstracts ($M = 247.87$, $SD = 294.86$) than to AI-authored ones ($M = 210.50$, $SD = 228.27$). This difference was, however, not significant, $t(410.46) = 1.49$, $p = .138$.
Thus, knowing that a text originates from an AI model did not independently increase the extent of editing, once the source information was disclosed. Overall, disclosure mainly affects authors’ effort when the abstract is human-authored. In contrast, knowing that the abstract is AI-generated does not elicit additional editing propensity. We conjecture disclosure primarily acts as a social rather than a purely evaluative signal.

\subsubsection{Heterogeneity Effects:} 
We probe whether the treatment effects on edits vary across \emph{author characteristics} (gender, age, education, profession) and \emph{AI perceptions/usage} (e.g., trust in AI, reported use of generative AI, willingness to use AI).
% For each of the three focal contrasts—\textbf{(i)} Human-noInfo vs.\ Human-withInfo (disclosure effect for human text), \textbf{(ii)} AI-noInfo vs.\ AI-withInfo (disclosure effect for AI text), and \textbf{(iii)} Human-withInfo vs.\ AI-withInfo (source effect with disclosure)—we estimate interaction models of the form
% \[
% \text{EditDistance}_{i}=\alpha+\beta_1 \,\text{Treatment}_{i}+\beta_2 Z_{i}+\beta_3(\text{Treatment}_{i}\times Z_{i})+\varepsilon_{i},
% \]
% where \(Z_i\) is a single moderator entered one at a time. We fit OLS with \textbf{cluster-robust} standard errors at the level of the original abstract (edits clustered by baseline abstract; 45–90 clusters across contrasts), using z-scored continuous moderators and dummy-coded categorical moderators. The term of interest is the interaction \(\beta_3\), which tests whether the treatment effect differs by \(Z\).
For Human-noInfo vs.\ Human-withInfo (\(N{=}444\)), none of the demographic or AI-perception covariates showed a reliable interaction with disclosure conditions.
For AI-noInfo vs.\ AI-withInfo (\(N{=}447\)), we observe a significant moderation by \emph{author's willingness to use AI}. Specifically, participants who reported \textit{they would use AI to edit/generate} exhibited a lower editing tendency ($p=0.030$) compared when authors who would not (see \autoref{fig:study2results}, Middle). 
% edited  a \emph{reduced} disclosure effect relative to the reference category: \(\beta{=}{-}184.91\), \(p{=}0.030\) for the interaction term \(\text{Treatment}\times\texttt{Would\_you\_use\_GPT{=}No}\).
% Intuitively, among authors who are averse to using GPT themselves, revealing AI origin does not further reduce the extent of editing as much as it does for the reference group (i.e., the disclosure effect is attenuated).
For Human-withInfo vs.\ AI-withInfo (\(N{=}441\)), two AI-perception moderators significantly differentiate the source effect. First, authors who believe \emph{others would use GPT} show a larger human-versus-AI difference \(\beta{=}+193.86\), \(p{=}0.005\).
Second, authors who reported \emph{they would not use GPT} exhibit a smaller human-versus-AI difference \(\beta{=}{-}189.80\), \(p{=}0.028\).
This implies that, when the source is always disclosed, these authors amplify their relative editing tendency on human-written abstracts versus AI-generated abstracts. On the other hand, individual reluctance toward AI is associated with a diminished human-vs.-AI gap in editing.
Overall, heterogeneity emerges primarily along perceptions about AI, not along demographics. Authors’ own willingness to use AI and their beliefs about others’ AI usage reliably mediate (i) how much AI disclosure affects editing of AI abstracts, and (ii) how strongly authors differentiate between human and AI abstracts when disclosure is present.

\subsubsection{Weak treatment effect in Confidence}
We examine two measures of authors' self-reported confidence: (a) \textit{original confidence}---the expected acceptance of the unedited abstract, and (b) \textit{confidence change}---the difference between confidence in the edited and original abstract.  
For each of the three pairwise treatment comparisons, we compute Welch’s unequal-variance \textit{t}-tests and cluster-robust OLS (clustered by provided abstracts). We then run exploratory mediation analyses for any effect achieving significance \(p < .10\).
Authors’ confidence does not differ significantly between the Human-noInfo and Human-withInfo conditions. Mean original confidence and post-edit confidence were statistically indistinguishable (\(p > .10\)), suggesting that disclosing human authorship does not influence how strongly authors believe the abstract would be accepted.
For AI-authored abstracts, a modest disclosure effect emerges. Authors show slightly higher \textit{original confidence} when AI authorship remains undisclosed (\(\beta \approx -0.15\), \(p = .08\)), as well as confidence change did not differ significantly. Thus, disclosure of AI authorship marginally lowers authors’ original confidence in the abstract’s acceptability, though it did not change their post-editing confidence gain.
With disclosure (Human-withInfo vs.\ AI-withInfo), neither original confidence nor confidence change differed significantly (\(p > .10\)), indicating that once source information is provided, perceived human or AI authorship alone does not affect confidence.
Given the marginal effect of AI-disclosure on original confidence, we examine whether perceived readability or beliefs on AI mediate the effect.  
% Among continuous candidates—\textit{perceived readability} (\texttt{readability\_reported}), \textit{AI trust} (\texttt{trust AI}), and \textit{belief about AI use prevalence} (\texttt{belief\_use\_genAI\_writing})—only \textbf{perceived readability} showed a meaningful path structure.  
We find disclosure of AI authorship slightly reduces perceived readability (\(p = .06\)), and readability positively predicts original confidence (\(p = .02\)).  
The Sobel test for the indirect effect remains marginal (\(z = 1.72, p = .085\)). Lower perceived readability may partially mediate the small decline in original confidence when AI authorship is disclosed (see \autoref{fig:study2results}, Right).  
% The results imply that authors’ initial belief in the acceptability of an abstract is somewhat sensitive to AI disclosure—showing a mild confidence penalty for AI-labeled text—but not for human-labeled text.  
This reduction primarily manifests through \textit{perceived readability}: when an abstract is identified as AI-written, authors perceive it as slightly less readable, which in turn lowers their confidence.
Post-editing confidence gains, however, remain stable across treatments, indicating that authors regain confidence after their edits.

\begin{figure}[t!]
  \centering
  \vspace{-5pt}
  \includegraphics[width=0.8\columnwidth]{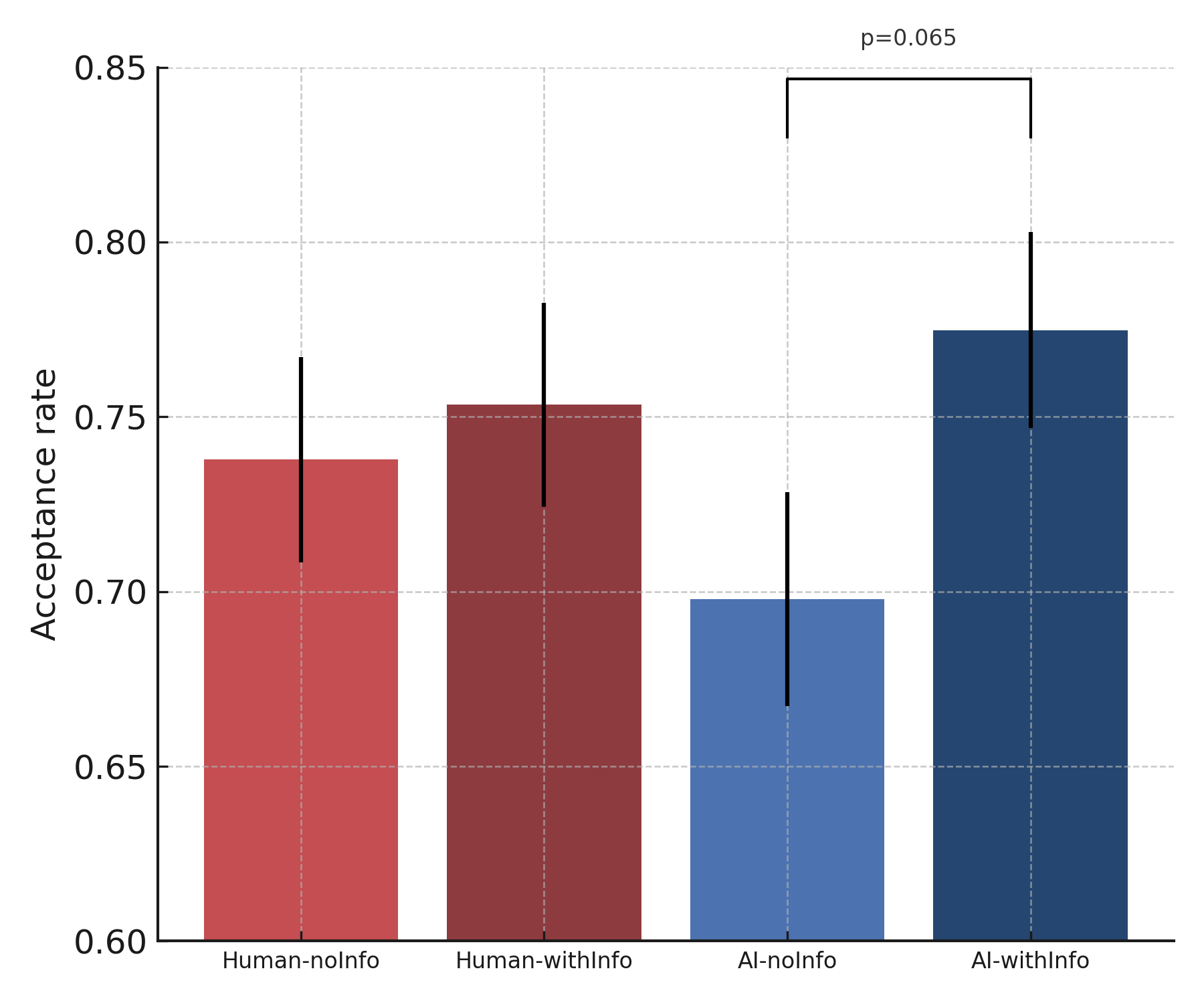}
  \vspace{-7pt}
  \caption{Reviewer decisions differ in AI conditions}
  \label{fig:study2_acceptance}
  \vspace{-7pt} % tighten space below caption
\end{figure}

\subsubsection{Reviewer decisions}
We next examine whether the acceptance rate by the reviewers differs across treatment conditions.
We model the final decision as a binary variable (\(1=\)Accept, \(0=\)Reject) and compare acceptance rates across the three treatment contrasts, using both $2\times2$ proportion tests and cluster-robust linear probability models with standard errors clustered by the original abstract. 
The acceptance rates remain nearly identical, with no statistical significance, across edited abstracts from human treatments: 0.730 in Human-noInfo versus 0.736 in Human-withInfo (\(\Delta{=}{-}.006\)).
For edited abstracts from AI treatments, acceptance rates are 0.706 in AI-noInfo versus 0.773 in AI-withInfo, indicating a modest increase under disclosure (\(\Delta{=}{-}.068\), \(p{=}0.081\)). This suggests that edited versions of AI-generated abstracts, when authors know about the source, make definitive editing decisions that lead to higher acceptance, as shown in  \autoref{fig:study2_acceptance}.

Keeping disclosure effect constant, comparing acceptance of edited abstracts from Human-withInfo (0.736) versus AI-withInfo (0.773) yields no significant difference. 
% Once source was disclosed, reviewers did not appear to systematically favor human or AI-authored texts.
% Across all three contrasts, the probability that an edited abstract was accepted remained largely stable, with only one marginal tendency: \emph{disclosing AI authorship slightly improved acceptance odds}. This effect aligns directionally with earlier findings on authors’ behavior—where disclosure influenced editing magnitude—but here it manifests in reviewers’ evaluative outcomes.
We further investigate how AI disclosure to authors may increase acceptance. Focusing on the marginal contrast in acceptance in edited abstracts from (AI-noInfo vs.\ AI-withInfo), we examine potential mediators reflecting editing effort, confidence, perceived readability, and AI beliefs. 
For each candidate mediator \(M\), we estimate (i) \(M{\sim}\text{Treatment}\) and (ii) \(\text{FinalDecision}{\sim}\text{Treatment}+M\) using cluster-robust SEs by baseline abstract; the indirect effect was tested via Sobel’s \(z\).
Among all mediators, editing magnitude through edit distance exhibits the strongest pattern. Disclosure of AI authorship slightly shifts the volume of edits (\(p{<}.10\)), and higher editing effort predicts a higher likelihood of acceptance (\(p{<}.05\)), producing a marginally significant indirect effect. 
% None of the other mediators—\texttt{confidence\_original}, \texttt{confidence\_edited}, \(\Delta\)\texttt{confidence}, \texttt{readability\_reported}, \texttt{trust AI}, \texttt{belief\_use\_genAI\_writing}, or \texttt{Can distinguish?}—showed reliable indirect effects. 
% This indicates that disclosure of AI authorship may subtly raise the probability of an edited abstract being accepted, and this effect appears to be partly channeled through the nature of authors’ revisions. 
Authors make slightly different editing decisions when aware of AI authorship, and these editorial behaviors modestly enhance the final review outcome.

\section{Study C: Quantitative and Qualitative Analyses of Edits}
\label{sec:studyc}

% \sk{should we mention any of these findings in the intro or abstract?}
Stylistic guidance for scientific writing is grounded in empirical work on how readers cognitively process text. Psycholinguistics research shows that comprehension improves when information arrives in a predictable fashion. Readability of scientific writing improves when the topic position for context is known to the readers or the stress position for new or important information is expected \cite{gopen1990science,halliday1967notes}. 
Clarity in scientific prose improves when grammatical subjects mention real ``characters'' and verbs express actions, rather than nominalizing them \cite{swales2014genre,williams2010style}. We perform post-hoc quantitative and qualitative analyses of the edits to elicit editing strategies emerging from different treatment conditions.

% At the discourse level, scholars show that coherence and persuasion depend on recognizable rhetorical moves. Swales’s Create-A-Research-Space model reveals that effective introductions first establish shared context, then identify a gap, and finally announce the study’s claim (Swales, 1990). Hyland’s corpus analyses add that stance markers like “we show” guide disciplinary readers through argument and position the author within a community (Hyland, 2004). Style handbooks synthesize these findings into layered advice: link sentences through lexical echo for cohesion, vary sentence length for rhythm, and end with “heavy” words—nouns or nominalizations—to maximize emphasis (Williams & Bizup, 2016; Sword, 2012). Thus, contemporary stylistic prescriptions are empirical, growing from evidence about reader expectations, genre conventions, and cognitive load rather than from taste alone.

\subsection{Quantitative Stylistic Metrics for Better Scientific Writing}

% To translate these reader-oriented principles into something we can test at scale, we selected one computational proxy from each of the course’s major style concepts—Action, Characters, Cohesion & Coherence, Emphasis, and Sentence Shape. Nominalization Density captures the Action mandate to keep verbs alive; Passive-Voice Share operationalises the Characters advice to foreground real agents; Topic-Continuity Index represents Cohesion & Coherence by tracking whether successive sentences stay on the same topic; New-at-End Score models the Emphasis guideline to deliver new information in the stress position; and Subject-Onset Distance, Initial-Momentum, and Pre-Main-Clause Length jointly cover the Sentence Shape goal of reaching the subject and verb quickly. Each metric can be computed with nothing more than a part-of-speech tagger and first-order dependency labels (e.g., nsubj, ROOT, nsubjpass), making the framework fast enough to run on thousands of abstracts while remaining transparent and interpretable for writers and reviewers alike.

Building on the reader-centric principles of clarity for scientific writing, we operationalize 36 computational linguistic metrics encompassing broad themes of writing: Action, Characters, Cohesion \& Coherence, Emphasis, and Sentence Shape. Each metric can be computed with a part-of-speech tagger and first-order dependency labels (e.g., nsubj, ROOT, nsubjpass), making them tractable for post-hoc analysis. We use Spacy\footnote{\url{https://spacy.io/}} to compute these linguistic features at scale for all provided and edited abstracts to understand how authors edit to improve clarity in their submissions. In subsequent analysis, we find seven of these features stand out that explain the most variance in the data and show significant trends across treatments. Each of these metrics quantify a cognitive signal through abstract's surface form: 1) \textit{Nominalization Density} detects when actions are hidden in nouns; 2) \textit{Passive-Voice Share} reveals disguised agency of the main sentence character; 3) \textit{Topic-Continuity Index} represents cohesion \& coherence by tracking whether successive sentences maintains the same topic; 4) \textit{New-at-End Score} tests whether novel information is delivered in the stress position; and 5) \textit{Subject-Onset Distance}, 6) \textit{Initial-Momentum}, and 7) \textit{Pre-Main-Clause Length} jointly estimates the sentence shape objective of reaching the subject and verb quickly. All metrics except Topic-Continuity and New-at-End Score trending high are a favorable sign of clear scientific writing.

\subsubsection{Stylistic Shifts Across Treatments}

We analyze how authors changed the linguistic structure of abstracts across four conditions.  
For each of 36 stylistic metrics, we computed the change from the original to the edited version (\(\Delta = \text{edited} - \text{original}\)) and compared conditions using Welch’s unequal-variance \(t\)-tests. Within each comparison, we controlled for multiple testing.  
% Effect sizes are reported as Cohen’s \(d\) based on pooled standard deviations.

Across all conditions, there was no reliable stylistic effect of disclosure alone.  
% Comparing authors who saw disclosure versus those who did not (within the same text source) yielded no significant differences after Holm adjustment (all \(p_{\text{Holm}} > .10\); median \(|d| = 0.10\)).  
This indicates that simply knowing whether the text was AI- or Human-generated did not meaningfully change how participants edited its stylistic form.

\begin{figure*}[t!]
  \centering
  % Left plot
  \begin{minipage}[t]{0.48\linewidth}
    \centering
    \includegraphics[width=\linewidth]{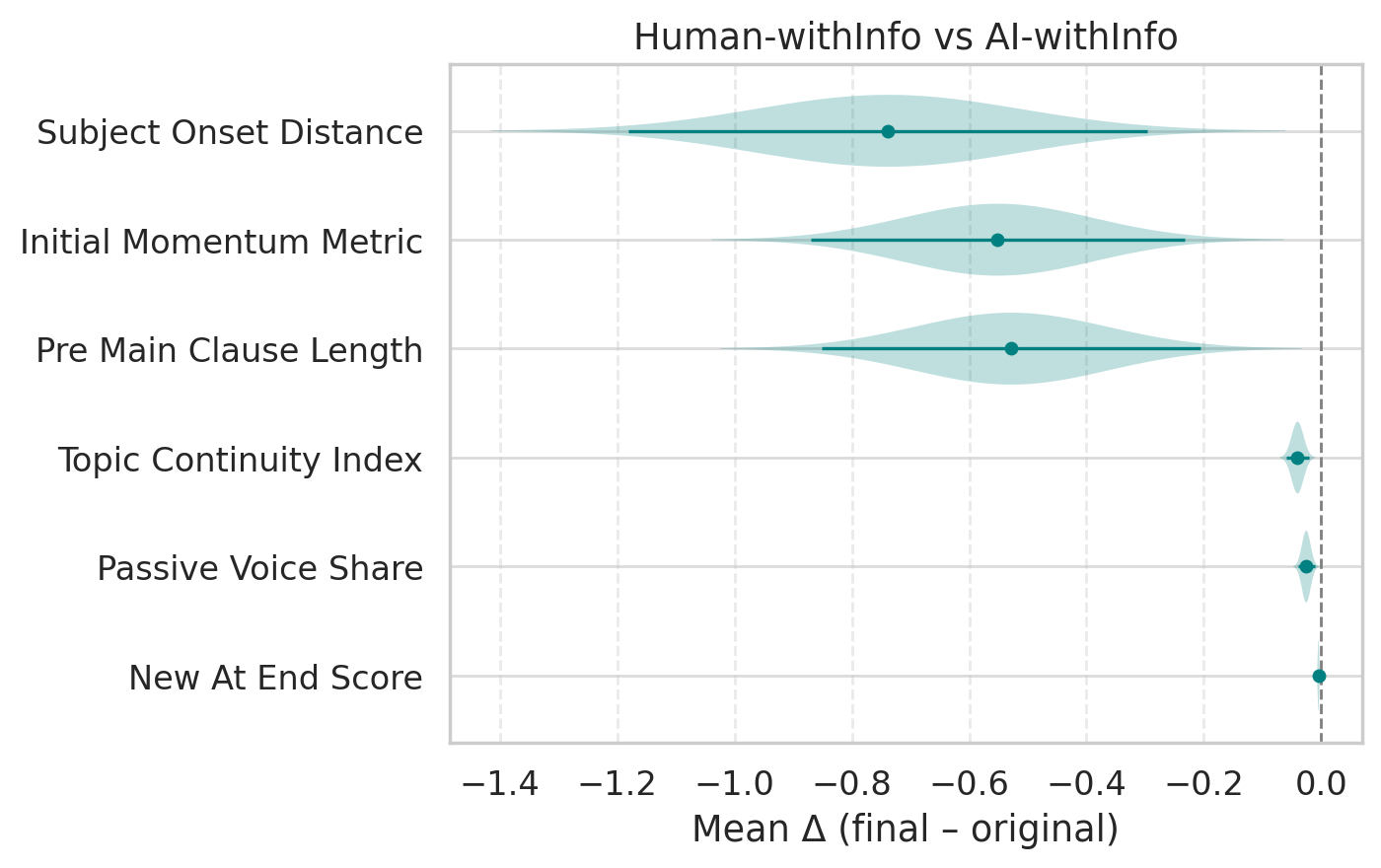}
    % \caption*{\textbf{Human-noInfo vs AI-noInfo}}
  \end{minipage}
  \hfill
  % Right plot
  \begin{minipage}[t]{0.48\linewidth}
    \centering
    \includegraphics[width=\linewidth]{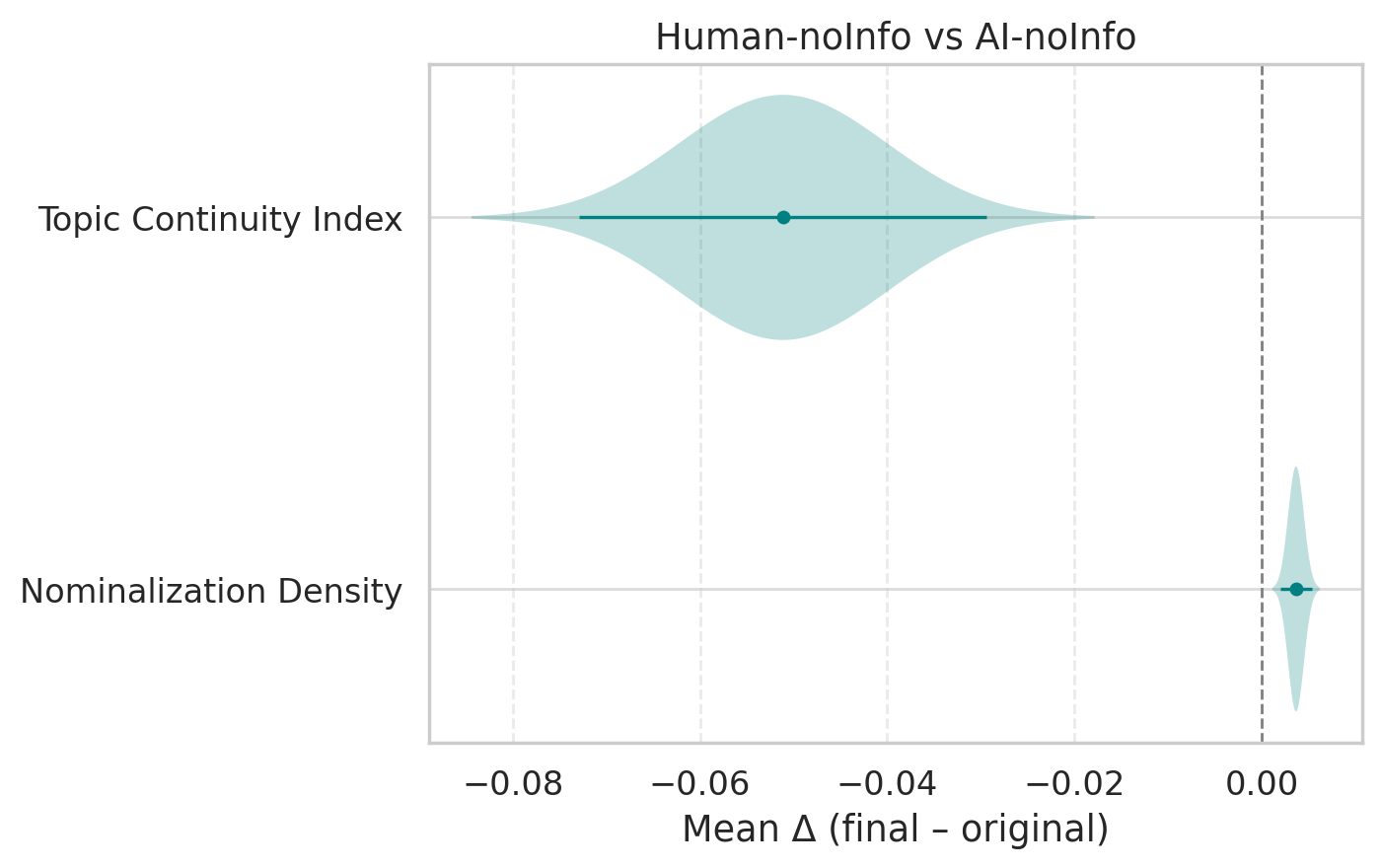}
    % \caption*{\textbf{Human-withInfo vs AI-withInfo}}
  \end{minipage}
  \vspace{-0.5em}
  \caption{Mean stylistic differences ($\Delta$) between edited and original abstracts across conditions.
  Shaded bands denote uncertainty distributions around 95\% CIs.}
  \label{fig:forest_side_by_side}
  \vspace{-1em}
\end{figure*}

By contrast, several stylistic differences emerged based on the source of the abstract being edited, as shown in \autoref{fig:forest_side_by_side}.  
When editing AI-generated versus Human-generated abstracts without disclosure (\textit{Human-noInfo vs.\ AI-noInfo}), authors produced lower Topic-Continuity Index (\(\Delta_{\text{AI}} - \Delta_{\text{Human}} = +0.051\); \(p=0.0002\)) and higher Nominalization Density (\(\Delta_{\text{AI}} - \Delta_{\text{Human}} = -0.0036\); \(p=0.0007\)). These effects suggest that authors, when revising human-written texts without source information, tend to make them read heavier and slightly less cohesive, whereas edits to AI-generated abstracts achieve a smoother topical cohesion and fewer nominalizations.

Noteworthy stylistic changes appear when authors edited in disclosure conditions (\textit{Human-withInfo vs.\ AI-withInfo}).  
All six metrics show significant changes:  
Topic-Continuity Index (\(p=0.0032\)),  
New-at-End-Score (\(p=0.0032\)),  
Passive-Voice Share (\(p=0.021\)),  
Initial-Momentum (\(p=0.024\)),  
Subject-Onset Distance (\(p=0.036\)), and  
Pre-Main-Clause Length (\(p=0.044\)).  
Across all six, edits to AI-generated abstracts were to make them structurally shorter, less passive, and more information-dense, while edits to human-written abstracts were to induce longer openings and delayed sentence subjects.  
% These effects demonstrate that stylistic conciseness and syntactic efficiency were primarily shaped by the \textbf{origin of the text}, not by whether that origin was disclosed.

% \paragraph{Interpretation.}  

\subsection{Edits and AI-detection}

To measure the significance of the edits and how AI being the source (perceived or disclosed) moderates editing behavior, we use Pangram\footnote{\url{https://www.pangram.com/}}, an automated AI detection tool, to find cases where author's edit made the edited abstract more human-written. We can only analyze cases where the provided abstract was AI-generated (i.e., AI-noInfo and AI-withInfo) and use AI detection to label if the edited abstract still can be perceived as ``AI-generated.''

We first examine treatment-level differences in the probability that an edited abstract is no longer classified as AI. Authors edited abstracts in the disclosure condition (AI-withInfo) to make the abstract more human-like (23\% cases, $N = 225$), i.e., can no longer be classified as AI-generated, as compared to no-disclosure condition (AI-noInfo; 18\% cases, $N = 225$), although this difference is not statistically significant. To characterize this behavior through character-level edits, we find that disclosure significantly increases editing intensity, making edited abstracts more human-like. Conclusively, character-level edit distance emerges as a strong predictor of AI detection outcomes ($p < 0.001$).

% To probe underlying mechanisms, we next examine the role of textual distance between the original and edited abstracts (dist). Controlling for treatment and education, dist is a strong and statistically significant predictor of detector outcomes: abstracts that differ more substantially from their original versions are markedly more likely to evade AI detection. Importantly, including dist as a control does not materially alter the treatment comparison or reveal heterogeneous disclosure effects, indicating that variation in detector classification is primarily driven by the extent of textual modification rather than differential responses to disclosure across participant characteristics.}

\subsection{Post-experiment Author Interviews}

To align our quantitative understanding of how authors edit several treatment conditions, we conducted a series of semi-structured interviews following the main experiment. Studies 1 and 2 quantify how the implicit source and its disclosure moderate editing behavior, whereas these interviews attempt to elicit the reasoning behind such editing behavior. 
% The interviews allowed us to probe how participants perceived quality, ownership, and fairness in AI-assisted editing, and how they made sense of their own decisions when the source of text was revealed or concealed. 
We interviewed 6 authors across all four treatment conditions who successfully completed the editing task. Each interview, conducted over Zoom, lasted for 15 minutes, for which they were compensated with a \$10 flat fee. Post-hoc thematic analysis of the transcripts reveals how authors adopted several editing strategies and their perception of AI-assistance in scientific writing.

% Interviews lasted 30–45 minutes, were conducted over Zoom, and were transcribed verbatim. Using reflexive thematic analysis, we identified recurring patterns in how participants described their editing strategies, perceptions of AI-generated text, and reflections on disclosure.

% To 
% Our qualitative interviews (six participants across four treatment conditions) reveal how editors approached scientific‐abstract editing under differing source and disclosure conditions.  
% Across the dataset, editing was not a mechanical correction process but a reflective act of sense-making and judgment.  
% We identified six interrelated themes that capture how participants framed, executed, and rationalized their edits.

\paragraph{Improving readability}
Authors across all conditions described a common goal of ``making the abstract easier to read.''  However, their editing strategies were different depending on treatments. When editing AI‐generated abstracts, authors simplified language and syntax, whereas when editing human‐written abstracts, they reorganized structure and emphasis. One AI–NoInfo editor explained, \emph{``I was just thinking how to make it easier for someone else to understand, but still having all the details''} (P3).  
% Another (P6) summarized, \emph{``I was trying to make the abstract more simple even for people who are not familiar with the topic.''}  
In contrast, authors in Human conditions focused on structure rather than verbosity: \emph{``I broke it into paragraphs to make it easier to read''} (P1).  
% Simplification thus took different forms—linguistic streamlining for AI texts, structural streamlining for human texts.

\paragraph{Restructuring for emphasis}
Beyond clarity, authors re-ordered information to highlight what they saw as the ``core'' contribution. Those in the Human–withInfo condition displayed the strongest reordering edits, often bringing the paper’s novelty to the beginning.  
P5 mentioned: \emph{``Instead of giving the full background first, I wanted it to say what is being proposed right off the bat,''}.  
AI–WithInfo editors, by contrast, focused on organization: \emph{``It would be more legible if restructured into paragraphs that are logically organized''} (P4).  
No-information editors rarely altered structure, limiting their work to grammar or tense adjustments (P0).  
Disclosure of source thus encouraged rhetorical restructuring rather than surface-level edits.

\paragraph{Perception of authority}
Knowing who (or what) authored the abstract reshaped editors’ stance toward authority. Healthy skepticism exists for both types of abstracts; \emph{``Even if it’s written by experts, they can also make mistakes''} (P5) or \emph{``I always have the habit of double-checking because LLMs hallucinate''} (P4).
Human–WithInfo participants knew experts had written the originals, yet still intervened confidently: \emph{``Even if it’s written by experts, they can also make mistakes''} (P5).  
By contrast, authors in no disclosure conditions edited at face value, relying only on textual cues (P0).  

% AI–WithInfo editors drew on prior experience with generative systems to justify vigilant checking: \emph{``I always have the habit of double checking because LLMs hallucinate''} (P4).  
% By contrast, those without disclosure edited at face value, relying only on textual cues (P0).  
% Disclosure thus activated epistemic vigilance---editors scrutinized the text more, not less, when authorship was transparent.

\paragraph{Accountability}
Across treatments, authors assumed their role to uphold the scientific integrity of the abstracts.  
They exerted equal oversight to ensure accuracy and coherence, irrespective of human or AI-generated abstracts. 
% and human authors alike as competent but fallible partners, requiring human oversight to ensure accuracy and coherence.  
As P1 noted, \emph{``AI is like an employee—it needs to prove itself before I stop rechecking everything''}.  
Disclosure amplified the sense of accountability: \emph{``I still double check anything from AI,''} (P4).
% reflecting a broader pattern of conscientious verification.  
The editing exercise reveals the author's intrinsic nature of executing expertise and moral obligation to balance the efficacy and efficiency (by using AI) in the critical act of scientific communication.

\section{Discussion}

Here we summarize our key findings connecting studies A (\Cref{sec:studya}), B (\Cref{sec:studyb}), and C (\Cref{sec:studyc}), grounding in prior work and with possible design implications.
\vspace{-0.5em}
{\paragraph{\textbf{Perceived or disclosed source acts as a reputational signal.}}
Study A shows fewer edits to AI-generated abstracts when the source is undisclosed. 
% (Section 4.1.1, Fig. 5 Left, pp. 13–14).
Study B shows disclosure (weakly) reverses this pattern, especially reducing edits to human-written abstracts.
% (Section 5.1.1, Fig. 6 Left, pp. 15–16).
% Reviewer acceptance remains unchanged across source conditions (Sections 4.1.5, 5.1.4, Fig. 7, pp. 15–17).
% \textbf{Disclosure primarily operates as a social and reputational signal, not an informational one.}
This result is consistent with prior work showing that source labels activate algorithm aversion and source-identity bias even when content quality is held constant \citep{DZINDOLET2003697}.
% Disclosure significantly reduces edits to human-authored abstracts (Human-noInfo vs. Human-withInfo; Section 5.1.1, Fig. 6 Left, p. 15).
% No comparable independent increase in editing of AI-generated abstracts upon disclosure (Section 5.1.1, p. 15).
Authors explicitly reference deference, accountability, and authority in the interviews. 
% (Section 6.2, “Perception of authority,” pp. 19–20).
This aligns with evidence in prior work that perceived authorship and authority trigger deference and reduced intervention in evaluative and collaborative settings \citep{zhu2025humanbiasfaceai,FIEDLER2025100321}.
Collaborative systems that foster AI assistance in scientific writing then must reveal origin of text (e.g., through watermarking) that can assign the appropriate accountability of the authors. 

\vspace{-0.5em}
\paragraph{\textbf{Readability and stylistic edits on AI writing.}}
Authors express higher perceived readability for AI-generated abstracts in no-disclosure conditions.
% (Section 4.1.4, Fig. 5 Right, p. 14).
Still, they make meaningful stylistic edits such as reducing nominalization and improving cohesion. 
% (Section 6.1.1, Fig. 8, pp. 18–19).
Interview evidence additionally confirms simplification strategies for AI text, also tying back to perceived or disclosed source, acting as a reputational signal.
% and restructuring for human text (Section 6.2, pp. 19–20). 
These readability and stylistic patterns reflect broader findings from recent corpus studies showing that AI-generated texts often exhibit distinct linguistic profiles such as more uniform sentence structure and enhanced surface-level coherence \citep{xu2025patterns}.
AI writing assistants could improve these obvious sloppiness to meaningfully contribute in academic writing. 

\vspace{-0.5em}
\paragraph{\textbf{Expertise mediates trust on AI writing.}}
Authors with PhD-level experience exhibit completely opposite editing behavior, editing AI-generated abstracts heavily.
% (Section 4.1.2, Fig. 5 Middle, p. 14).
Mixed-effects models show variance driven by author characteristics rather than abstract quality, which was corroborated by
% (Section 4.1.2, p. 13).
interview narratives reflecting a stronger critical posture among experienced authors. 
% (Section 6.2, pp. 19–20).
Prior studies similarly show that experts are less susceptible to fluency-driven trust in AI-generated text and are more likely to detect limitations or stylistic artifacts, whereas less experienced users rely more heavily on surface-level coherence as a quality signal \citep{10.1145/3613904.3642037,chakrabarty2024art}.

\vspace{-0.5em}
\paragraph{\textbf{Asymmetric trust on human and AI}}
Authors report equal responsibility in checking correctness across AI and human text, but more weights put when the text is AI-generated, which was revealed consistently during interviews. 
% (Section 6.2, “Accountability,” pp. 19–20).
Indeed, source disclosure influences how authors edit, hence their trust allocation. When the source is disclosed as AI, we find a strong heterogeneity: authors willing to use AI reduce edits to AI-generated abstracts. On the other hand, with the source being disclosed as human, it substantially reduces the overall edit for human-written abstracts. But, in general, even though editing behavior changes across source information conditions depending on the disclosure, it does not translate into systematic differences in reviewer acceptance. This divergence suggests that disclosure primarily affects authors’ perception of reputation or a manifestation of algorithm aversion, rather than the substantive quality of the final text.
% authors lowers their tendency to edit when most notably suppressing revisions to human-authored abstracts—it does not translate into systematic differences in reviewer acceptance. This divergence suggests that disclosure primarily affects authors’ social and reputational calculus rather than the substantive quality of the final text. (Sections 5.1.1, 5.1.4, pp. 15–17).}
% \textcolor{red}{\textbf{Overall, LLMs function best as efficient first-draft generators rather than autonomous scientific writers.}
% AI-generated abstracts reach comparable acceptance only after expert editing (Sections 4.1.5, 5.1.4, pp. 15–17).
% Stylistic analyses show systematic improvements introduced by human editors rather than originating from the model (Section 6.1, Fig. 8, pp. 18–19).
% %Synthesis echoed in the Discussion and Conclusion sections (Sections 7–8, pp. 20).
% }
\vspace{-0.5em}
\paragraph{\textbf{Limitations and future work}}
Like most experimental studies, our setup also has some limitations which could be addressed in a future work. Though elaborate, our setup is lab-based: a similar study could be done in a real conference setup, where we could observe author behavior with natural incentives.
Similarly, capturing editing behavior with real-time AI help, contrary to our offline setup where AI abstracts are already provided.
Additionally, confidence and readability plays an important role in describing author's editing behavior, but we do not incentivize to provide true beliefs to keep the study simple.
Finally, This work establishes a foundation for evaluating AI-generated scientific writing in realistic incentivized authorial and review settings. Future research should extend this methodology to other domains beyond computer science to understand domain-specific interactions with AI. Investigating collaborative workflows such as co-writing between multiple authors with AI support would uncover both potential and pitfalls in the use of AI assistance. Additionally, scaling to full-paper evaluations and incorporating multimodal inputs (e.g., tables, figures) would provide a more comprehensive assessment of LLM utility in scientific communication.
Finally, data created using such studies could also be used to train models to conduct edits to further improve text quality in the domain of science.

\section{Conclusion}

We offer a systematic evaluation of AI-assisted scientific writing through a randomized controlled trial simulating a peer review process. Our results show that authorial behavior is significantly influenced by both the implicit source of an abstract and the disclosure of it: authors made the most revisions to human-written texts when source information was withheld, whereas AI-generated abstracts received more extensive edits when attribution was disclosed. However, reviewer decisions were not significantly affected by the abstract’s source. Instead, acceptance outcomes were primarily driven by the degree of meaningful edits. Our analysis suggests that \textit{careful edits on the AI-generated abstract can yield higher, if not equal, acceptance as compared to abstracts that are human-authored}. 

% Our analysis suggests extensive editing does not reliably improve acceptance rates, and that LLM-generated abstracts, particularly when unattributed, are often of sufficient quality to be accepted with minimal revision.

Looking beyond a standalone experiment, longitudinal studies could help us understand how repeated use of AI-assisted writing affects authors' skill development, stylistic habits, and confidence over time, as well as how it reshapes norms around peer review and scientific communication. Finally, we raise an important point: ethical considerations around transparency, attribution, and responsible use of AI in research are understudied and can have welfare-altering impacts \citep{Hazra2025AISS}. Our study highlights the importance of academic experiences for authors, which play an important role in understanding and effectively utilizing the benefits of AI assistance. At a large scale, we must expand training programs, which are lacking, to inform novice researchers about the socially beneficial use of AI in science. Community-driven investigations of responsible use of AI in science thus must uncover downstream effects on scientific rigor, reproducibility, and community norms to optimally reap the democratic benefits of AI.

% Ultimately, such work can support the development of evidence-based guidelines for AI-assisted writing: covering best practices for attribution, oversight of AI contributions, domain-specific adaptation, and ethical integration, ensuring that AI enhances rather than diminishes clarity, accuracy, and integrity in scientific communication.

\section*{GenAI Usage Disclosure}
We used GPT-4o to generate AI-generated abstracts central to our experiments. Similarly, GPT-4o was used for assigning reviewers to relevant abstracts for review, again critical for our experiments. We use generative AI-based tools (e.g., Grammarly) to make grammatical checks and grammatical edits for our manuscript.

\begin{acks}
This material is based upon work supported by the Ai2 Faculty Research Award and the OSU Kirwan Institute. We are grateful to Tuhin Chakraborty, Joseph Chang, Pao Siangliulue, and Peter Clark for feedback on early drafts of this paper. We also thank the anonymous reviewers for their insightful comments. 
\end{acks}
% \subsection{Extra text}
% Together, these themes illustrate editing as an epistemic and relational act.  
% Without disclosure, editors corrected surface errors, guided by readability alone.  
% With disclosure, they engaged critically: restructuring, verifying, and reflecting on authorship and trust.  
% While human-source texts invited rhetorical refinement, AI-source texts elicited simplification and verification.  
% Across conditions, editors viewed themselves as human‐in-the-loop collaborators who ``trust but verify,'' transforming a proofreading task into a reflective practice of co-authorship.

% Taken together, these results show that disclosure had little effect on how human authors rewrote text at the stylistic level.  
% Editing patterns reflected the linguistic characteristics of the input draft—whether it was originally produced by a human or by an AI model—rather than the label attached to it.  
% While disclosure did not alter stylistic mechanics, the consistent differences between AI- and human-origin drafts highlight how model-generated writing induces distinct editing trajectories: authors streamline and condense AI text but expand and nominalize human text.  
% These findings imply that stylistic adaptation operates below the level of explicit disclosure awareness, underscoring the need for design interventions that account for implicit style transfer when humans edit AI-assisted content.

% \subsubsection{Incentives}

\bibliographystyle{ACM-Reference-Format}
\bibliography{ref}

% \clearpage 
% \newpage
\appendix
\section{Additional Tables and Figures}

% \Needspace{10\baselineskip}
% \section{Editing Interface}\label{app:interface}

\begin{figure}[h]
    \centering
    \includegraphics[width=\linewidth]{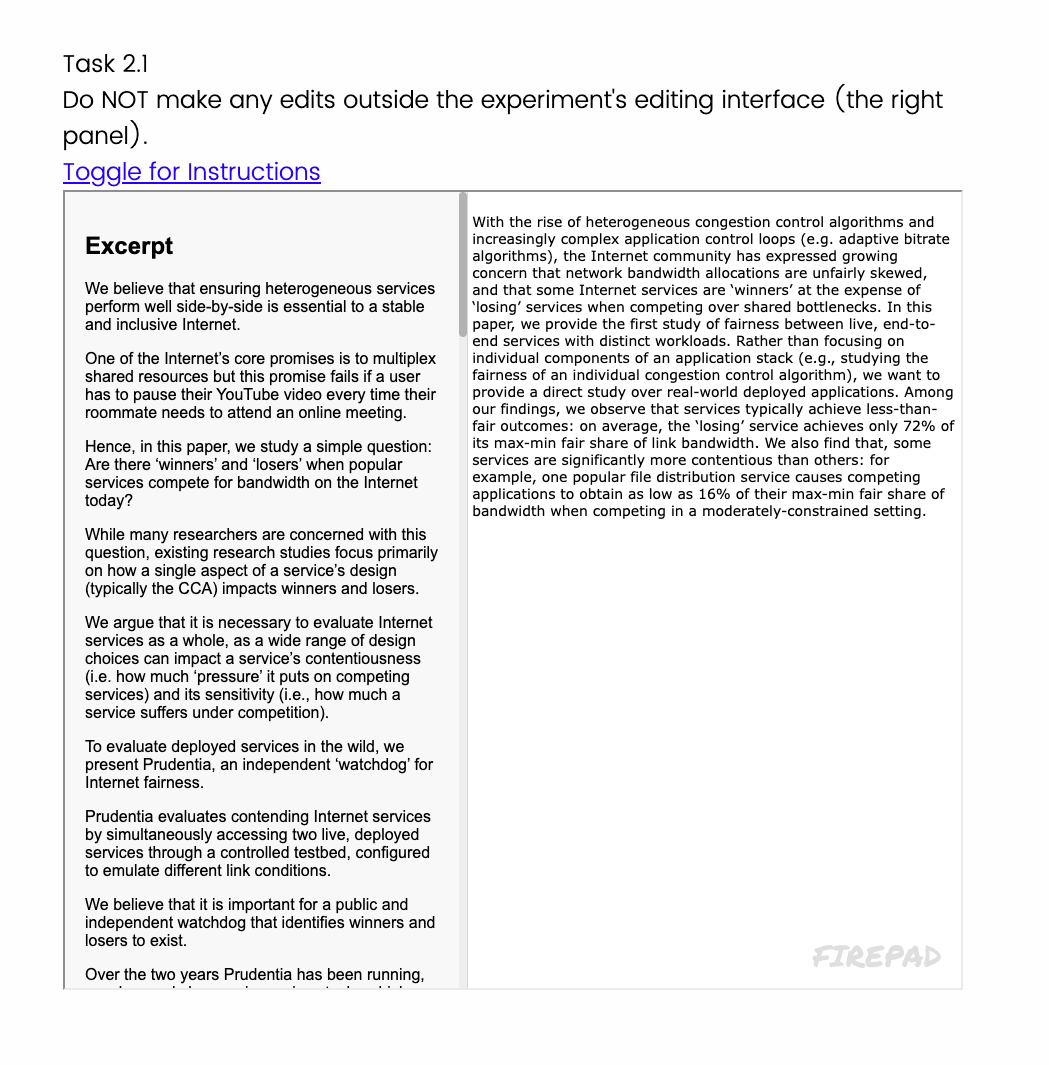}
    \vspace{-3em}
    \caption{This is how the authors view the abstracts. The left non-editable, copy-paste disabled panel includes the excerpts from a selected research paper, and the right panel is the embedded Firepad editor where the authors can make changes.}
    \label{fig:authorinterface}
\end{figure}

% \section{Screening Questions}\label{app:screening_questions}

\begin{figure}[h]
    \centering
    \includegraphics[width=0.9\linewidth]{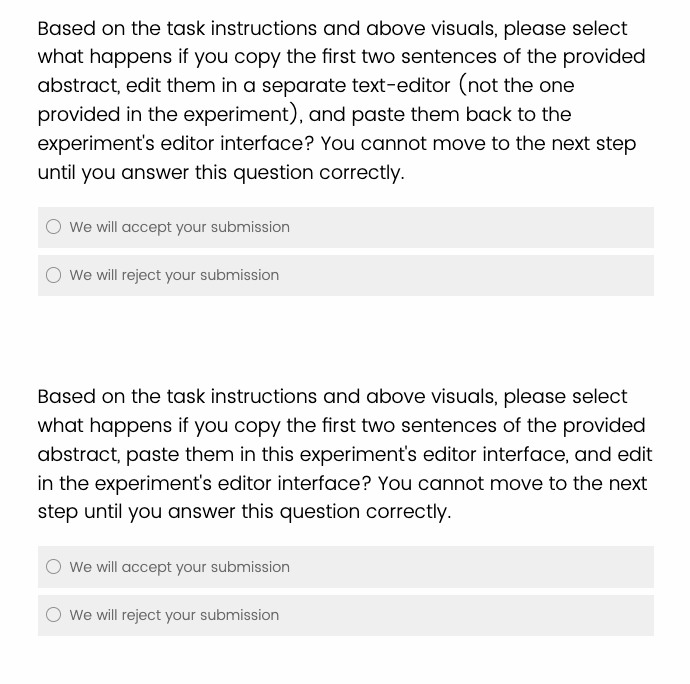}
    \vspace{-2em}
    \caption{Screening questions to assess the authors' understanding of the task.}
    \label{fig:screening_questions}
\end{figure}

\subsection{Author Comments}
We analyze the sentiment of the comments received by the authors. 139 authors out of 297 left a comment, and 54 left a comment indicating ``no comment''. Remaining 85 comments are divided into three topics: study (study design), abstract (experiences with the abstracts provided), and AI (opinions regarding the AI usage trend). \autoref{tab:commentsentiment} shows the sentiment analysis of each topic. Most reviewers are positive toward the study design, have a less positive opinion toward the abstracts, and are split about AI usage. Word cloud of authors' comments is in \autoref{fig:comments_author}.

\begin{table}[h!]
\centering
\caption{Author comment sentiment analyzed.}
\label{tab:commentsentiment}
\begin{tabular}{lrrrrl}
\hline
Topic & Positive & Negative & Positive Rate  \\
\hline
Study    & 62 & 9 & 0.873 \\
Abstract & 7  & 3 & 0.700 \\
AI       & 2  & 2 &  0.500 \\
\hline
\end{tabular}
\end{table}

\begin{figure}[h]
    \centering
    \includegraphics[width=0.8\linewidth]{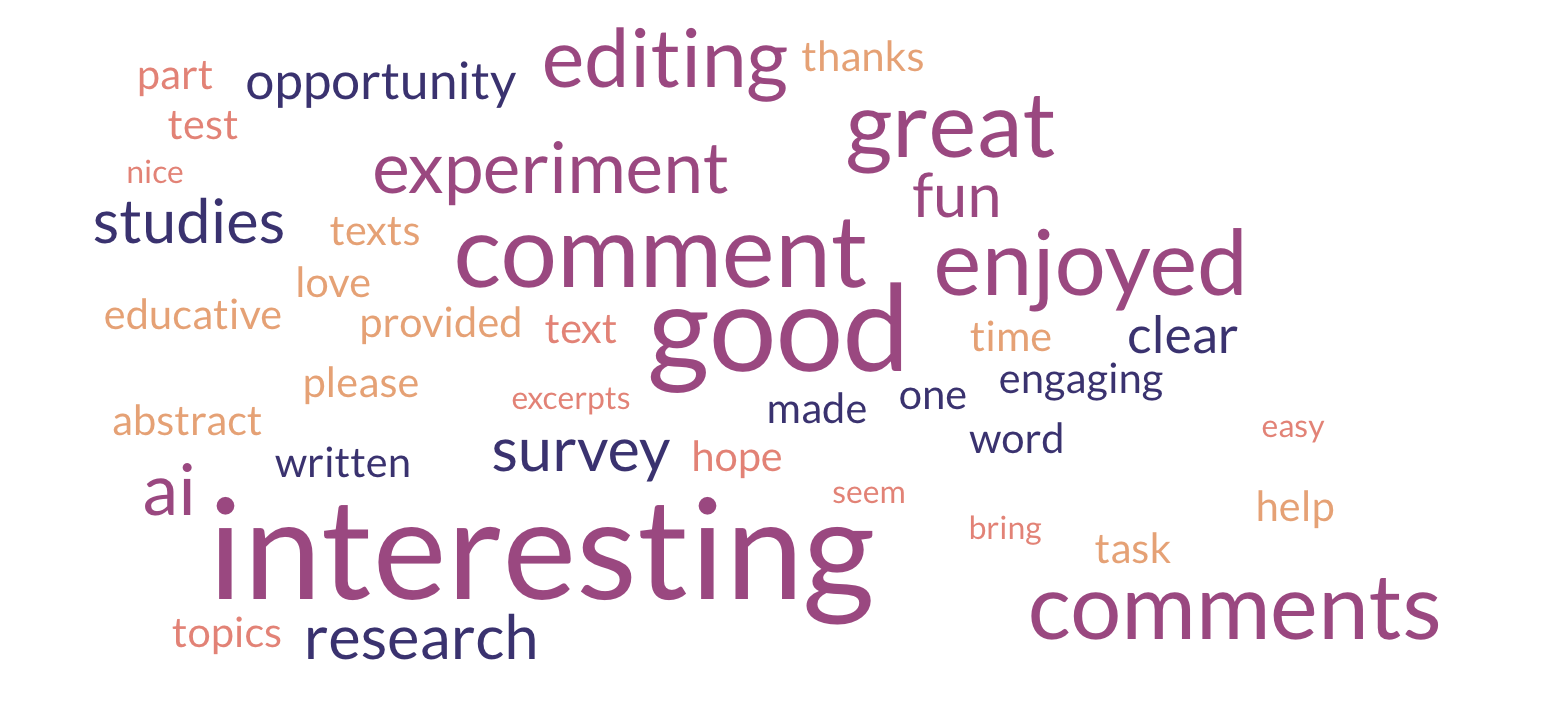}
    \caption{Word cloud for authors' comments.}
    \label{fig:comments_author}
\end{figure}

\subsection{Distinguishing AI and Human Texts}
We asked the same question to both authors and reviewers: ``Can you describe some distinguishing features that help you differentiate between AI-generated and human-written texts?'' Popular responses included differences in word choices, sentence structure, and tone of presentation. Some reported that they are barely distinguishable.
Word cloud for this question is in \autoref{fig:reason_distinguish}.

\begin{figure}[h!]
    \centering
    \includegraphics[width=0.8\linewidth]{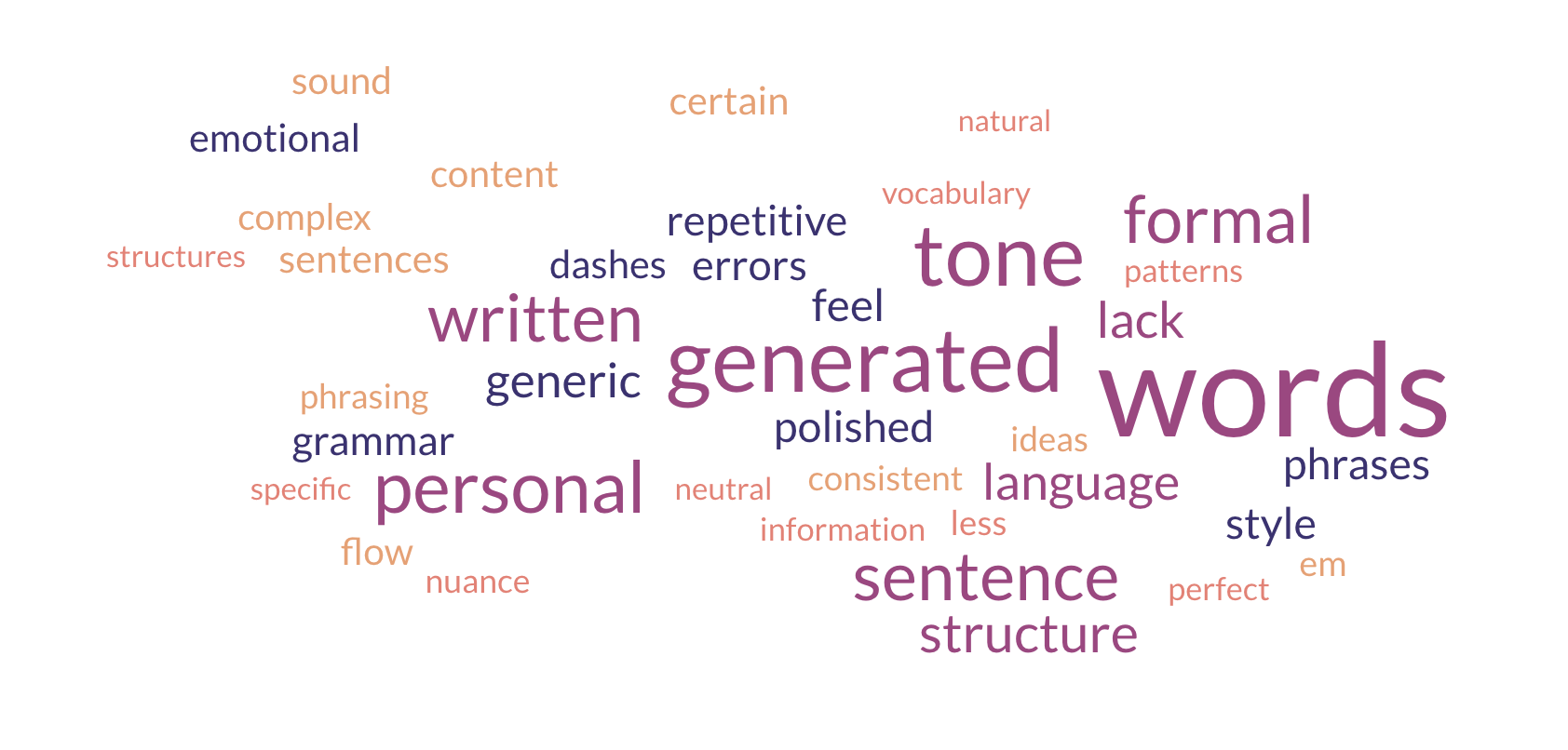}
    \caption{Word cloud for distinguishing factors between AI and human texts, from both authors and reviewers.}
    \label{fig:reason_distinguish}
\end{figure}

\begin{table*}[t]
\caption{\small List of selected papers with venues \& years}
\label{tab:paperlist}
\small
\setlength{\tabcolsep}{3pt}
\renewcommand{\arraystretch}{1}

% Global formatting for p-columns
\raggedright
\sloppy

\begin{tabular}{@{}
  p{0.75\linewidth}
  p{1.5cm}
  p{1cm}
  p{0.7cm}
@{}}
\toprule
Title & Venue & Citation & Year \\
\midrule
An Analysis of Robustness of Non-Lipschitz Networks & NeurIPS & 16 & 2024 \\
RedCode: Risky Code Execution and Generation Benchmark for Code Agents & NeurIPS & 37 & 2024 \\
InstructVideo: Instructing Video Diffusion Models with Human Feedback & CVPR & 62 & 2024 \\
Locally Adaptive Neural 3D Morphable Models & CVPR & 3 & 2024 \\
ToolSword: Unveiling Safety Issues of Large Language Models in Tool Learning Across Three Stages & ACL & 41 & 2024 \\
To be Continuous, or to be Discrete, Those are Bits of Questions & ACL & 4 & 2024 \\
Prudentia: Findings of an Internet Fairness Watchdog & SIGCOMM & 5 & 2024 \\
Rodeo: Making Refinements for Diverse Top-k Queries & VLDB & 2 & 2024 \\
Titan: Efficient Multi-target Directed Greybox Fuzzing & IEEE S\&P & 20 & 2024 \\
C-FRAME: Characterizing and measuring in-the-wild CAPTCHA attacks & IEEE S\&P & 3 & 2024 \\
An Architecture For Edge Networking Services & SIGCOMM & 5 & 2024 \\
ModsNet: Performance-aware Top-k Model Search using Exemplar Datasets & VLDB & 2 & 2024 \\
Learning-based Widget Matching for Migrating GUI Test Cases & ICSE & 29 & 2024 \\
Harp: Leveraging Quasi-Sequential Characteristics to Accelerate Sequence-to-Graph Mapping of Long Reads & ASPLOS & 4 & 2024 \\
GraphTrail: Translating GNN Predictions into Human-Interpretable Logical Rules & NeurIPS & 7 & 2024 \\
RainbowCake: Mitigating Cold-starts in Serverless with Layer-wise Container Caching and Sharing & ASPLOS & 50 & 2024 \\
Mining Pull Requests to Detect Process Anomalies in Open Source Software Development & ICSE & 1 & 2024 \\
Uncertainty-aware Action Decoupling Transformer for Action Anticipation & CVPR & 17 & 2024 \\
DocLLM: A Layout-Aware Generative Language Model for Multimodal Document Understanding & ACL & 91 & 2024 \\
More is Merrier: Relax the Non-Collusion Assumption in Multi-Server PIR & IEEE S\&P & 8 & 2024 \\
An exabyte a day: throughput-oriented, large scale, managed data transfers with Effingo & SIGCOMM & 3 & 2024 \\
FedSQ: A Secure System for Federated Vector Similarity Queries & VLDB & 5 & 2024 \\
Formal Mechanised Semantics of CHERI C: Capabilities, Undefined Behaviour, and Provenance & ASPLOS & 11 & 2024 \\
Code Impact Beyond Disciplinary Boundaries: Constructing A Multidisciplinary Dependency Graph and Analyzing Cross-Boundary Impact & ICSE & 3 & 2024 \\
Optimistic Critic Reconstruction and Constrained Fine-Tuning for General Offline-to-Online RL & NeurIPS & 1 & 2024 \\
DREAM: Diffusion Rectification and Estimation-Adaptive Models & CVPR & 8 & 2024 \\
Pareto Optimal Learning for Estimating Large Language Model Errors & ACL & 5 & 2024 \\
Distribution Preserving Backdoor Attack in Self-supervised Learning & IEEE S\&P & 37 & 2024 \\
NegotiaToR: Towards A Simple Yet Effective On-demand Reconfigurable Datacenter Network & SIGCOMM & 13 & 2024 \\
Composable Data Management: An Execution Overview & VLDB & 3 & 2024 \\
WASP: Workload-Aware Self-Replicating Page-Tables for NUMA Servers & ASPLOS & 8 & 2024 \\
Predicting Performance and Accuracy of Mixed-Precision Programs for Precision Tuning & ICSE & 11 & 2024 \\
DiffuserLite: Towards Real-time Diffusion Planning & NeurIPS & 28 & 2024 \\
DeCoTR: Enhancing Depth Completion with 2D and 3D Attentions & CVPR & 7 & 2024 \\
Hard Prompts Made Interpretable: Sparse Entropy Regularization for Prompt Tuning with RL & ACL & 4 & 2024 \\
eAUDIT: A Fast, Scalable and Deployable Audit Data Collection System & IEEE S\&P & 31 & 2024 \\
Fast, Scalable, and Accurate Rate Limiter for RDMA NICs & SIGCOMM & 6 & 2024 \\
SpannerLib: Embedding Declarative Information Extraction in an Imperative Workflow & VLDB & 1 & 2024 \\
ACES: Accelerating Sparse Matrix Multiplication with Adaptive Execution Flow and Concurrency-Aware Cache Optimizations & ASPLOS & 4 & 2024 \\
ACAV: A Framework for Automatic Causality Analysis in Autonomous Vehicle Accident Recordings & ICSE & 10 & 2024 \\
ChaosBench: A Multi-Channel, Physics-Based Benchmark for Subseasonal-to-Seasonal Climate Prediction & NeurIPS & 24 & 2024 \\
ViLa-MIL: Dual-scale Vision-Language Multiple Instance Learning for Whole Slide Image Classification & CVPR & 48 & 2024 \\
Generative Cross-Modal Retrieval: Memorizing Images in Multimodal Language Models for Retrieval and Beyond & ACL & 37 & 2024 \\
SoK: The Long Journey of Exploiting and Defending the Legacy of King Harald Bluetooth & IEEE S\&P & 20 & 2024 \\
OptimusPrime: Unleash Dataplane Programmability through a Transformable Architecture & SIGCOMM & 3 & 2024 \\
\bottomrule
\end{tabular}
\end{table*}

\clearpage

\begin{figure*}[h]
    \centering
    \includegraphics[width=\linewidth]{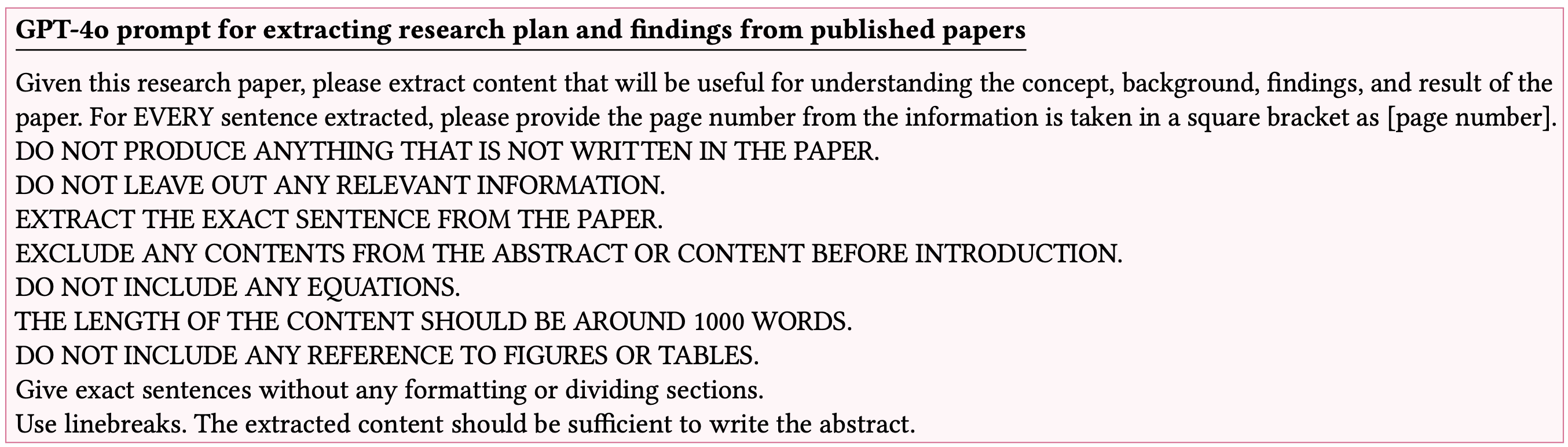}
    \caption{GPT-4o Prompt we used to extract research findings from papers. The output of the model (after verification) is used as the excerpt to generate the abstract.}
    \label{fig:promptforfindings}
\end{figure*}

\begin{figure*}[h]
    \centering
    \includegraphics[width=\linewidth]{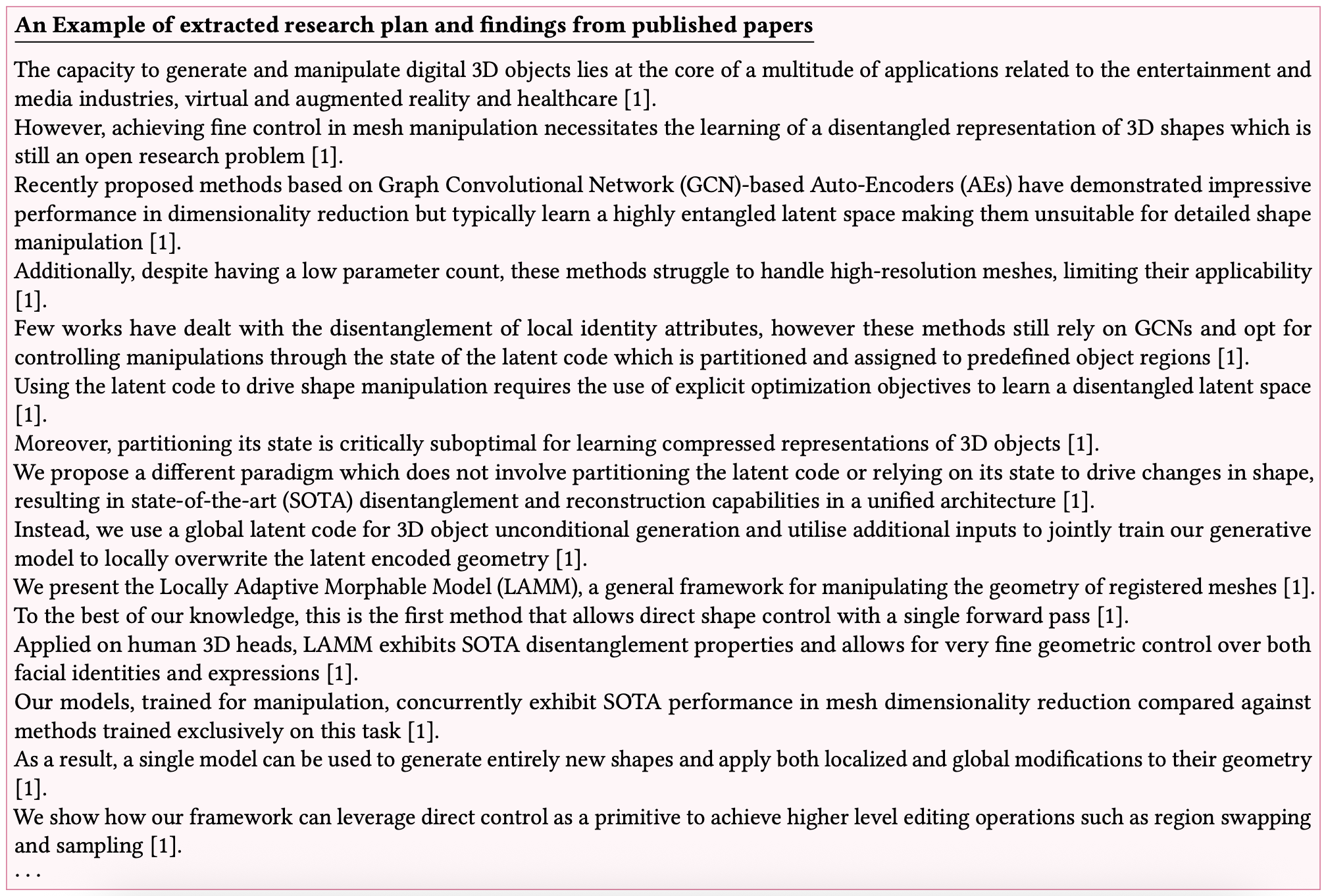}
    \caption{An example of research findings fed as input to LLMs to generate abstracts. It is generated using the paper ``Locally Adaptive Neural 3D Morphable Models'' published at CVPR.}\label{fig:sampleresearchfindings}
\end{figure*}

\begin{figure*}[h]
    \centering
    \includegraphics[width=\linewidth]{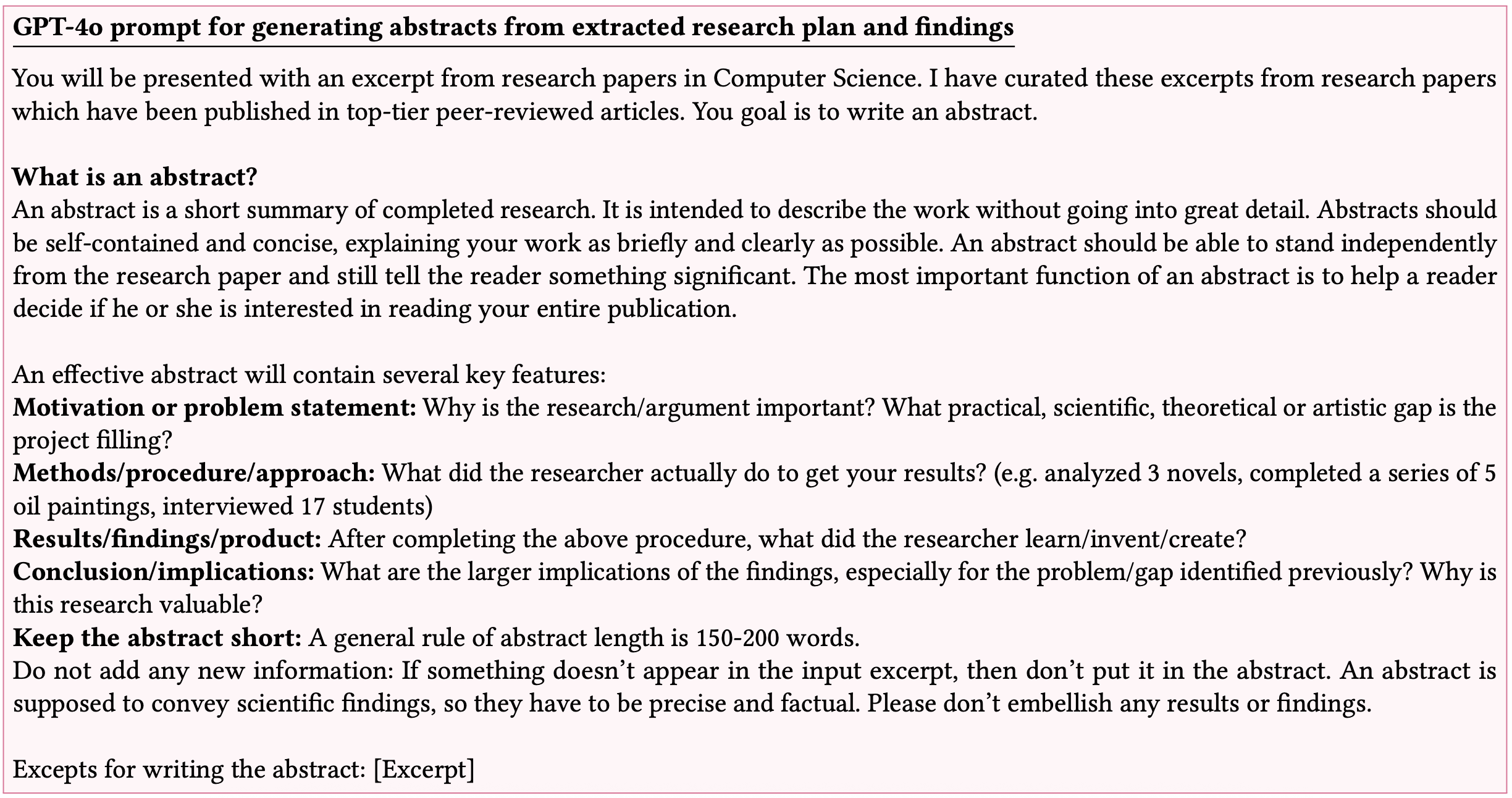}
    \caption{GPT-4o Prompt we used to generate abstract given research findings (excerpts).}\label{fig:promptforabstract}
\end{figure*}

\begin{figure*}[h]
    \centering
    \includegraphics[width=\linewidth]{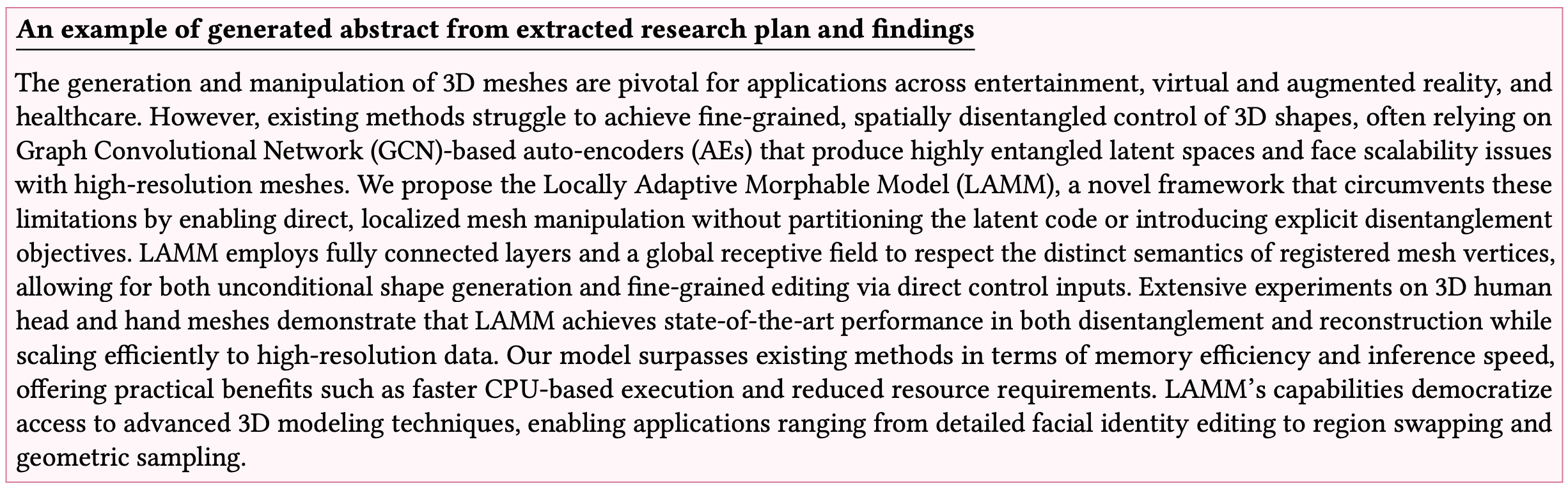}
    \caption{An example of AI-generated abstract.}\label{fig:sampleabstract}
\end{figure*}

\begin{figure*}[h]
    \centering
    \includegraphics[width=\linewidth]{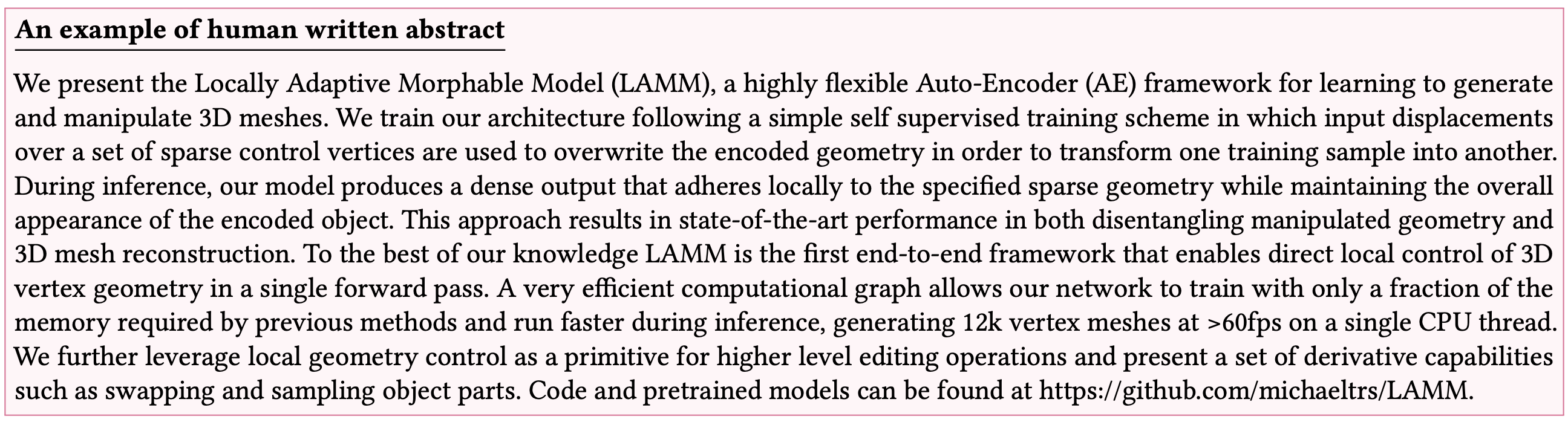}
    \caption{An example of human-written abstract from the published papers.}\label{fig:sampleabstract_human}
\end{figure*}

\end{document}